%% file: manuscript.tex
\documentclass[12pt]{article}
\usepackage[utf8]{inputenc}  % already the default in Overleaf, but make it explicit
\usepackage[T1]{fontenc}     % better font rendering, especially for accented letters
\usepackage{amsthm,amsmath,amssymb,amsfonts,bm,enumerate,xcolor,graphicx,natbib,color,url}
\RequirePackage[colorlinks,citecolor=blue,urlcolor=blue]{hyperref}
\usepackage{booktabs}
\usepackage{multirow}  % Add this line
\usepackage{hyperref}
\usepackage{subcaption} 
\usepackage{afterpage}
\usepackage{ulem}
\usepackage{pifont}
\usepackage{textcmds}
\usepackage{booktabs}
\usepackage{multirow}
\usepackage{algpseudocode}
\usepackage{booktabs}
\usepackage{adjustbox}
\usepackage[usestackEOL]{stackengine}
\usepackage[toc,page]{appendix} 
\usepackage{algorithm}
\usepackage{lipsum}
\usepackage{setspace}
\usepackage{geometry}
\usepackage{siunitx}
\usepackage{subcaption,dcolumn}
\usepackage{setspace}
\allowdisplaybreaks

\definecolor{lb}{RGB}{44,139,183}

\usepackage{ulem}

\input{macros.tex}

\newcommand{\blind}{1}
\renewcommand{\arraystretch}{2}

\colorlet{shadecolor}{gray!40}

\begin{document}
\thispagestyle{empty}
\baselineskip=28pt
\vskip 5mm

\begin{center}{\Large{\bf {\color{black}{Scalable Spatiotemporal Modeling for Bicycle Count Prediction}}}}

\end{center}
\baselineskip=12pt

\vskip 2mm

 \if1\blind
{
\begin{center}
\large
Rishikesh Yadav$^1$, Alexandra M. Schmidt$^2$, Aur\'elie Labbe$^1$,  Pratheepa Jeganathan$^3$, Luis F. Miranda-Moreno$^4$\\
\end{center}
\footnotetext[1]{
\baselineskip=10pt Department of Decision Sciences, HEC Montréal, Canada. Emails: rishikesh@iitmandi.ac.in; aurelie.labbe@hec.ca}
\footnotetext[2]{
\baselineskip=10pt Department of Epidemiology, Biostatistics, and Occupational Health, McGill University, Canada. E-mail: alexandra.schmidt@mcgill.ca}
\footnotetext[3]{Department of Mathematics and Statistics, McMaster University, Canada. Email: jeganp1@mcmaster.ca }
\footnotetext[4]{Department of Civil Engineering, McGill University, Canada. Email: luis.miranda-moreno@mcgill.ca}
 } \fi

\baselineskip=26pt
\vskip 2mm
%\centerline{\today}
%\vskip 4mm
\begin{center}
    {\large{\bf Abstract}} 
\end{center}
\baselineskip=26pt
We propose a novel sparse spatiotemporal dynamic generalized linear model for efficient inference and prediction of bicycle count data. Assuming Poisson distributed counts with spacetime-varying rates, we model the log-rate using spatiotemporal intercepts, dynamic temporal covariates, and site-specific effects additively. Spatiotemporal dependence is modeled using a spacetime-varying intercept that evolves smoothly over time with spatially correlated errors, and coefficients of some temporal covariates including seasonal harmonics also evolve dynamically over time. Inference is performed following the Bayesian paradigm, and uncertainty quantification is naturally accounted for when predicting bicycle counts for unobserved locations and future times of interest. To address the challenges of high-dimensional inference of spatiotemporal data in a Bayesian setting, we develop a customized hybrid Markov Chain Monte Carlo (MCMC) algorithm. To address the computational burden of dense covariance matrices, we extend our framework to high-dimensional spatial settings using the sparse SPDE approach of Lindgren et al. (2011), demonstrating its accuracy and scalability on both synthetic data and Montreal Island bicycle datasets. The proposed approach naturally provides missing value imputations, kriging, future forecasting, spatiotemporal predictions, and inference of model components. Moreover, it provides ways to predict average annual daily bicycles (AADB), a key metric often sought when designing bicycle networks. 
\noindent  


\baselineskip=26pt 
\section{Introduction} 

\subsection{General}
Understanding and modeling bicycle count data over space and time is crucial for improving urban mobility \citep{zhou2015understanding, chen2021dynamic} and optimizing transportation systems \citep{liu2020development, ahmadi2024optimizing}, particularly as cities increasingly adopt bicycle-sharing programs \citep{fishman2016bikeshare}  to promote sustainable transport \citep{teixeira2021empirical}, enhance infrastructure planning \citep{teixeira2020does}, optimize resource allocation \citep{ricci2015bike}, and improve the biking experience \citep{sersli2022easy}. Bicycle count data are often collected by automated sensors, which are the most common and efficient techniques for collecting bike data that uses inductive loop sensors (buried in the ground) and infrared sensors (placed on posts) to detect and quantify bicycle as they traverse the area. 

Modeling bicycle demands present severeal challenges with several factor affecting the demand \citep{eren2020review}. For example, these data are inherently count-based and often follow non-Gaussian distributions such as Poisson or Negative Binomial  that complicates traditional statistical modeling. Additionally, the data exhibit strong non-stationarity, with substantial temporal variations; due to seasonal cycles, weekday-weekend effects, and long-term trends \citep{zhao2016modeling}; as well as spatial heterogeneity across different locations. Furthermore, spatiotemporal dependencies are common, where bicycle activity at one station influences patterns at nearby stations. Another challenge is missing data, often caused by sensor malfunctions or incomplete observations, which can significantly affect the accuracy of prediction. This issue is further compounded by the high dimensionality of bicycle count data, as it involves data from multiple monitoring stations (sensors) with high-frequency temporal resolution, leading to large, complex datasets that require scalable and computationally efficient models. In addition to observed bicycle colunts, predictive accuracy also depends on additional covariates, such as weather conditions, special events, and infrastructure developments. While incorporating these factors can enhance model performance, it also introduces further complexity, as their effects may vary across space and time, requiring a flexible and interpretable modeling approaches to capture these dependencies effectively. 

To address these challenges,  we propose a spatiotemporal dynamic generalized linear modeling (ST--DGLM) framework that integrates dynamic generalized linear models \citep{west1985dynamic,west1997bayesian} with classical spatiotemporal statistical methods \citep{Banerjee.etal:2014}.  Each model component is designed to capture the underlying pattern in the data, such as a random walk for time trends, harmonics of appropriate order to capture seasonality, appropriate design matrices to capture the effects of exogenous covariates, and the use of specialized random effects to capture spatiotemporal dependence. The proposed approach follows a full Bayesian inference framework, providing natural uncertainty quantification while maintaining interpretability and strong predictive performance. To ensure computational feasibility for large-scale applications, we adopt a sparse representation of precision matrices of spatial random effects via Gaussian Markov random fields (GMRF), using the stochastic partial differential equation (SPDE) approach of \cite{lindgren2011explicit}.  

A key application of our modeling framework is the estimation of average annual daily bicyclists \citep[AADB,][]{nordback2013estimating, kaiser2025counting}, a widely used metric in cycling studies. Automated bicycle counting systems typically consist of two types of monitoring sites: long-term sites, where permanent counters (sensors) record data for long-term, and short-term sites, where sensors are temporarily placed for limited durations. AADB at long-term sites is straightforward to compute by averaging daily counts over the year, but estimating AADB for short-term sites is more challenging, as it often requires interpolation using nearby long-term site data. Standard interpolation techniques introduce errors due to their inability to fully account for temporal fluctuations, seasonal variations, and other non-stationarities in the data \citep{beitel2017quality}. The proposed approach {\it explicitly} incorporates these factors through a Bayesian hierarchical approach, improving AADB estimation and naturally providing uncertainty quantification.

\subsection{Related Work}  
\label{subsec:Intro_relatedwork}  
Most existing models for bicycle counts rely mainly on available covariates and do not explicitly account for spatiotemporal dependencies. These approaches often include regression-based models such as generalized linear models (GLMs) \citep{ermagun2018bicycle, hochmair2019estimating} and Bayesian regression models \citep{strauss2013cyclist}. While simple, effective, and interpretable in predicting and estimating bicycle counts, these models often struggle to capture the complex nonlinear relationships between counts and predictors affecting uncertainty quantification of predictions. In more complex settings, recent advancements in machine learning (ML) and deep learning (DL) have been increasingly applied to transportation data modeling, particularly for bicycle volume prediction. 
For instance, \cite{sekula2018estimating} used fully connected artificial neural networks (ANNs) to estimate hourly traffic volumes, while \cite{das2020interpretable} applied machine learning methods such as Random Forest (RF), Support Vector Machine (SVM), and K-Nearest Neighbor (KNN) to low-volume roadway traffic estimation. Similarly, \cite{islam2016estimation} employed deep neural networks (DNNs) and SVMs to estimate average annual daily traffic (AADT). More recently, \cite{feng2024spatial} developed a spatial-temporal aggregated graph neural network for forecasting docked bike-sharing demand, and \cite{zhou2024bike} introduced a spatiotemporal graph convolutional network for modeling bike-sharing usage. In principle there is a whole plethora of machine and deep learning methods that are applied to bicycle demand and forecasting; see \cite{albuquerque2021machine, miah2023estimation} for an overview of widely used ML/DL models for predicting bicycle counts and AADB estimation. While ML/DL models are highly effective in capturing complex nonlinearities, their limited interpretability makes them less suitable for transportation planning and decision-making. Furthermore, these models often lack straightforward methods for uncertainty quantification, typically relying on bootstrapping or ensemble techniques. Motivated by these limitations, the models developed in this paper aim to strike a balance between interpretability and prediction performances; i.e., they combine the interpretability of simple GLM-based approaches with the flexibility to capture nonlinear relationships and achieve predictive performance comparable to that of ML/DL models, all while providing natural uncertainty estimates within the Bayesian paradigm.

A third class of models consists of statistically-driven spatiotemporal methods, which define explicit spatiotemporal dependencies and offer a balance between interpretability and flexibility. These models capture spatiotemporal correlations through structured random effects and have been successfully applied in various studies to capture spatiotemporal dependencies \citep{Banerjee.etal:2014}. However, such models remain underutilized in bicycle demand modeling, with most studies still relying on regression-based techniques or ML approaches. However, some studies have explored such spatiotemporal modeling in transportation contexts for predicting shared bicycle demand. For example, \cite{guidon2020expanding} compared spatial regression models with random forests and found that incorporating spatial dependencies improved predictive performance; \cite{almannaa2020dynamic} applied dynamic linear models for modeling site-specific demand variations over time; \cite{duan2021applying} developed a spatiotemporal model using Integrated Nested Laplace Approximations (INLA) to incorporate spatial and temporal dependencies; and more recently, \cite{lei2022bayesian, lei2024scalable, lanthierbktr} proposed factorized low-rank spatiotemporal models to handle missing data and make large-scale predictions for transportation data efficiently, including the bicycle volume predictions.  

Additionally, time-varying parameter models, such as state-space models  \citep{lee2008missing, goh2012online, wang2016improved, song2016deeptransport} have been widely used in transportation studies and could be adapted for modeling bicycle counts data as well. Dynamic linear models \citep[DLMs,][]{west1997bayesian, petris2009dynamic} are well-known state-space models; known for their ability to capture intricate temporal non-stationarity within data, efficiently modeling temporal trends, seasonality, and the effects of exogenous covariates, all of which are typical characteristics of  bicycle count data. \cite{pherwani2024spatiotemporal} extends such models by proposing dynamic generalized linear models for each site separately, providing a framework for future forecasting. These models may be further adapted for spatiotemporal modeling by incorporating spatial random effects. For instance \cite{buddhavarapu2021new} proposed a spatiotemporal model for crash frequency data, incorporating intrinsic conditional autoregressive \citep[ICAR,][]{besag1995conditional} priors to capture spatial dependencies and a DLM structure to account for temporal dependencies. 

The proposed framework builds on similar developments by incorporating a fully Bayesian hierarchical structure with explicit spatiotemporal dependencies. In contrast to \cite{buddhavarapu2021new}, which relies on ICAR priors, we adopt an SPDE-based Gaussian Markov Random Field (GMRF) framework \citep{lindgren2011explicit, rue2005gaussian}, which enables full-rank spatial dependencies, is graph-invariant, can be applied to point-referenced spatial locations, and offers scalable computational Bayesian inference methods. Furthermore, unlike \cite{buddhavarapu2021new} that focus solely on negative binomial models, our framework may be generalized to multiple GLM families, making it applicable to a broader range of transportation datasets.

\subsection{Main Contributions}
The proposed spatiotemporal modeling framework includes specific modeling components designed to efficiently capture the complex structure in the bicycle data while maintaining prediction accuracy and interpretability. Specifically, we incorporate spatiotemporally varying intercepts, temporally varying coefficients for some important covariates, seasonal harmonics, and exogenous covariates to model the desired components effectively that can be further extended to include large spatial dimensions. 
Furthermore, if the bicycle count follows a non-Gaussian distribution, such as Poisson, we cannot derive closed-form full conditionals for Bayesian inference. 
To address this, we propose an efficient MCMC sampler that leverages gradient and Hessian information via the preconditioned Metropolis-adjusted Langevin algorithm  \citep[pMALA,][]{girolami2011riemann} for parameters lacking closed-form full conditionals. We demonstrate the scalability and effectiveness of the proposed MCMC sampler through a synthetic data experiment. Further, an application to Montreal’s Ecocounter bicycle data shows its comparable performance against standard Bayesian GLM, BKTR \citep{lanthierbktr}, and BayesNF \citep{saad2024scalable}. 
In summary, the key contributions of this paper include:
\begin{itemize}
\setlength{\itemsep}{-2mm} 
    \item A dynamic generalized spatiotemporal model enabling imputation, spatial interpolation, forecasting, spatiotemporal predictions; all along with a natural uncertainty estimaties.
    \item Direct estimation of AADBs at short-term and new spatial locations along with their uncertainty estimates.
    \item An extension to high-dimensional spatial data using sparse precision matrices exploiting the SPDE approach of \cite{lindgren2011explicit}.
    \item A unified framework and MCMC sampler for the inference and prediction of our proposed modeling framework, implemented in the \texttt{R} package \texttt{sparseDGLM} for broader transportation applications.
\end{itemize}

The remainder of this paper is structured as follows: Section \ref{sec:modelingframework} presents the proposed hierarchical model. Section \ref{sec:BayesInf_prediction} details Bayesian inference and prediction strategies. Section \ref{sec:synthetic_exp} discusses synthetic experiments validating the proposed MCMC sampler and showcasing its scalability. Section \ref{sec:application} applies the proposed model to the Eco-counter dataset, and Section \ref{sec:conclusion} summarizes the paper's  main findings and outlines for future research directions.

\section{Proposed model}
\label{sec:modelingframework}
In this section, we detail our proposed modeling framework, with general model details in Section \ref{subsec:dens-DGLM}, and in Section \ref{subsec:SPDEdetails}, we detail the sparse model counterparts using the SPDE approach that extends the general modeling framework to include a large number of spatial locations. 
\subsection{Spatiotemporal dynamic generalized linear model (ST--DGLM)}
\label{subsec:dens-DGLM}
Let $ Y_t(\bm{s}) $ denote a spatiotemporal counting process observed at spatial location $\bm s\in \mathcal{S} $ within a continuous spatial domain $\mathcal{S} \subset \mathbb{R}^2$ and discrete time (e.g., window of 15 minutes interval) $t \in \mathcal{T} = \{1,\ldots,\, T\}$. Our spatiotemporal dynamic generalized linear model (ST--DGLM) framework assumes that the counting process $Y_t(\bm s)$ follows independent Poisson distributions conditioned on the spatiotemporal varying log-rate parameter $\lambda_t(\bm{s})$, i.e.,
\begin{align}
Y_t(\bm{s}) \mid \lambda_t(\bm{s}) &\stackrel{\rm{ind}}{\sim} \mathrm{Poisson}\left[\exp\{\lambda_t(\bm{s})\}\right],\quad \bm s\in \mathcal{S} \subset \mathbb{R}^2, t \in \mathcal{T}. 
\end{align}
The spatiotemporal dependence is incorporated in the log-rate parameter by additively including fixed effects, a dynamic temporal structure, structured spatiotemporal random effects (here spatiotemporal intercepts) and random independent errors. Specifically,
\begin{align}
\label{eq:gen-st-model-intensity-level}
\lambda_t(\bm{s}) &= \mu_t(\bm s) + \bm{F}_t' \bm{\theta}_t + \bm{X}_t(\bm{s})'\bm{\beta} + \varepsilon_{t}(\bm{s}),
\end{align}
where $\mu_t(\bm{s})$ denotes a spacetime varying intercept or structured spatiotemporal random effects defined as in the traditional dynamic linear models (DLM) structure; see \cite{petris2009dynamic, schmidt2019dynamic}.  More explicitly,  
\begin{align}
\label{eq:gen-model-st-intcpt}
    \mu_t(\bm s) &= \mu_{t-1}(\bm s) + \omega_t(\bm s), \quad t \in \mathcal{T}, \quad  
    \mu_0(\bm s) = \mu_0,  \quad \text{mostly} \quad \mu_0 = 0,
\end{align}  
where the errors $\omega_t(\bm s)$ follow a Gaussian process with a Mat\'ern covariance \citep{guttorp2006studies} to capture spatial dependencies.  The Mat\'ern  covariance function for pair of spatial locations $\bm s$ and $\bm s'$ is defineds as:  
\begin{align}
\label{eq:matern}
\Sigma(\bm s, \bm s') &= \sigma^2 \Omega(\bm s, \bm s'), \quad  
\Omega(\bm s, \bm s')  = \frac{1}{\Gamma(\nu) 2^{\nu-1}} \left(\frac{d(\bm s, \bm s')}{\kappa}\right)^\nu K_{\nu}\left(\frac{d(\bm s, \bm s')}{\kappa}\right),
\end{align}  
where $\Omega(\bm s, \bm s')$ denotes a Mat\'ern correlation between $\bm s$ and $\bm s'$; $\sigma^2$, $\kappa$, and $\nu$ are the marginal variance, range, and smoothness parameters, respectively; $d(\bm s, \bm s')$ is the Euclidean distance between $\bm s$ and $\bm s'$; and $K_\nu(\cdot)$ is the modified Bessel function of  second kind.  

The component $\bm{F}_t' \bm{\theta}_t$ in \eqref{eq:gen-st-model-intensity-level} with $\bm{F}_t$ being a $p$-dimensional vector of purely temporal (spatially constant) covariate denotes the purely temporal component with a discrete-time dynamic structure defined as:
\begin{align}
\bm{\theta}_t &= \bm{G}_t \bm{\theta}_{t-1} + \bm{v}_t, \quad t \in \mathcal{T}, %\quad \bm{\theta}_0 \sim \mathcal{N}_p(\bm{m}_0, \bm{C}_0), 
\quad \bm{v}_t \sim \mathcal{N}_p(\bm 0, \bm{W}),  
\end{align}
where  $\bm{G}_t$ is a $p \times p$ state evolution matrix defining how the state vector evolves from time $t-1$ to time $t$, and $\bm{v}_t$ is the state noise, assumed to follow a multivariate normal distribution with mean zero and covariance matrix $\bm W_t$. For simplicity, we assume $\bm W_t = \bm W$ to be a diagonal matrix. The dynamics of $\theta$'s are completed by assuming that the initial state vector $\bm{\theta}_0$ follows an independent Gaussian distribution with known mean $\bm{m}_0$ and a diagonal covariance matrix $\bm{C}_0$ with large variances. 

Furthermore, $\bm{\beta}$ in \eqref{eq:gen-st-model-intensity-level} represents the covariate coefficients associated with the $q$ spatiotemporally varying covariates $\bm{X}_t(\bm{s}) = \{\bm{X}_t^1(\bm{s}), \ldots, \bm{X}_t^q(\bm{s})\}'$. The term $\varepsilon_{t}(\bm{s})$ follows an independent and identically distributed (i.i.d.) normal distribution with mean zero and variance $\tau^2$ that accounts for small-scale variation in the log--rate $\lambda_t(\bm{s})$ not explained by the spatial, temporal, and other model components. These random effects are beneficial to our modeling framework, as they simplify our proposed MCMC-based inference by enabling closed-form full conditional distributions for many model parameters; see, Appendix \ref{appd:appednA} and Section~S1 in the Supplementary Materials for more details. 

\subsection{Sparse ST--DGLM}
\label{subsec:SPDEdetails}
The spatially structured random effects (intercept) in \eqref{eq:gen-model-st-intcpt} are assumed to follow a multivariate Gaussian distribution with a Mat\'ern covariance matrix \eqref{eq:matern}, which is usually dense (containing many non-zero elements). Therefore, in high spatial dimensions, calculating the multivariate Gaussian likelihood becomes computationally intensive, making it challenging to fit models at large spatial scales \citep{heaton2019case, hazra2024exploring}. Specifically, computing the inverse and determinant of an $n \times n$ covariance matrix involved in the multivariate Gaussian density requires $\mathcal{O}(n^3)$ operations, a cost that grows substantially as the number of spatial locations increases. 
There have been several efforts in the statistical literature to address high-dimensional matrices in the multivariate Gaussian distribution, including approaches such as Vecchia approximations \citep{vecchia1988estimation, katzfuss2021general}, nearest neighbor Gaussian processes \citep{datta2016hierarchical}, block composite likelihood techniques \citep{eidsvik2014estimation}, and SPDE approaches \citep{lindgren2011explicit}. For a detailed review of these methods, see \cite{heaton2019case} and \cite{hazra2024exploring}. 
In practice, no single approach consistently outperforms others in terms of computational efficiency and accuracy, and selection often relies on cross-validation and experimentation. As an alternative to the ST-DGLM model described in Section \ref{subsec:dens-DGLM}, we propose a sparse version of ST-DGLM based on an SPDE-based approximation of the Mat\'ern field \eqref{eq:matern}. This approach is widely used and forms the basis of the \hyperlink{https://www.r-inla.org/}{\texttt{R-INLA}} software \citep{rue2009approximate, lindgren2015bayesian} for spatial modeling by efficiently handling sparse precision matrices instead of dense covariances.

For notational simplicity, we omit the time subscript \( t \) from the spatially structured random effects (intercept) \( \mu_t(\bm{s}) \) in \eqref{eq:gen-model-st-intcpt} and let $ \mu(\bm{s}) $ be a dense Gaussian process with a Mat\'ern correlation function defined as in \eqref{eq:matern}.  \cite{lindgren2011explicit} established a link that approximates the continuously indexed Gaussian random process $ \mu(\bm{s}) $ by a discrete GMRF \citep{rue2005gaussian},  obtained as the solution of the following SPDE:
\begin{equation}
(8 \kappa^{-2} - \Delta)^{\alpha/2}\sigma \mu(\bm s) = \mathcal{W}(\bm s), \quad \bm{s}\in \mathcal{S} \subset \mathbb{R}^2,
\label{eq:SPDEapprox}
\end{equation}
where $\Delta = \partial^2/\partial x^2 + \partial^2/\partial y^2$ is the Laplacian operator, $\kappa$ is the range parameter, $\alpha$ controls the smoothness of the realizations (related to $\nu = \alpha - d/2$ in \eqref{eq:matern}), $\sigma$ is related to marginal variance, and $\mathcal{W}(\bm s)$ is a Gaussian spatial white noise process. For dimension $d=2$, $\alpha=2$ is a natural choice \citep{whittle1954stationary}, which results in smoothness parameter $\nu=1$ in \eqref{eq:matern}. Since an exact solution is not possible for the SPDEs in \eqref{eq:SPDEapprox}, a numerical approximation based on finite element methods over a triangular mesh in $\mathbb{R}^2$ may be performed using the Delaunay triangulation \citep{lindgren2011explicit, lindgren2015bayesian}. More explicitly,
let $\mathcal{S}^{\star}=(\bm s_1^{\star},\ldots,\bm s_N^{\star})$ denote the mesh locations, 
then the finite elements representation of the solution of SPDEs \eqref{eq:SPDEapprox} may be written as $\mu(\bm s) = \bm{a}(\bm s)'\bm R$, where $\bm{a}(\bm s)= (a(\bm s, \bm s_1^{\star}), \ldots, a(\bm s, \bm s_N^{\star}))^{'}$ is an $N$-dimensional vector with $a(\bm s, \bm s_j^{\star})=1$ if the location $\bm s$ falls within a triangle with $\bm s^{\star}$ being one of the vertices, and 0 otherwise; for an example of such mesh, see Figure~\ref{fig:mesh-nodes_synth_exp}. The weight vector $\bm R=(R(\bm s_1^*),\ldots,R(\bm s_N^*))^{'}$ is defined at the mesh nodes $\mathcal{S}^\star$ and follows a Gaussian distribution with mean vector $\bm 0$ and sparse precision matrix $\bm Q$ defined as:
\begin{equation}
\bm{Q}= {{1}\over{\sigma^2}}\bm Q_{\kappa}, \quad \bm Q_{\kappa} = \frac{\kappa^2}{4\pi} \left( \frac{1}{\kappa^4} \bm{C} + \frac{2}{\kappa^2} \bm{G}_1 +  \bm{G}_2\right),
\label{eq:PrecisionSPDE}
\end{equation}
where $\bm C$, $\bm G_1$, and $\bm G_2$ are $N \times N$-dimensional sparse matrices that are completely determined by the construction of the mesh.
For more details, see \cite{lindgren2015bayesian, bakka2018spatial, cisneros2023combined}.  

For spatiotemporally varying $\mu_t(\bm s)$, we have corresponding space-time varying weights $R_t(\bm s^{\star})$ and hence $\bm R_t = (R_t(\bm s_1^\star),\ldots, R_t(\bm s_N^\star))'$. Therefore, we replace $\mu_t(\bm{s})$ in equation \eqref{eq:gen-st-model-intensity-level} with $\mu_t(\bm{s}) = \bm{a}(\bm{s})' \bm{R}_t = \sum_{k=1}^{N} a(\bm{s}, \bm{s}_k^*) R_{kt}$, such that the log-rate of the Poisson distribution for sparse ST-DGLM model is now given by:
\begin{align}
\lambda_t(\bm{s}) &=  \bm a(\bm s)' \bm R_t + \bm{F}_t' \bm{\theta}_t + \bm{X}_t(\bm{s})'\bm{\beta} + \varepsilon_{t}(\bm{s}),
\end{align}
where the spatiotemporal dependence is specified in the weights $R_t(\bm s^*)$; i.e.,
\begin{align*}
R_t(\bm s^*) &= R_{t-1}(\bm s^*) + \omega_t(\bm s^*), \quad t \in \mathcal{T}, \quad
R_0(\bm s^*) = 0 ,\\
\bm{\omega}_t &= \left\{\omega_t(\bm{s}_1^\star),\ldots, \omega_t(\bm s_N^\star)\right\}' \sim \mathcal{N}_N(\bm{0}, \sigma^2\bm Q_\kappa^{-1}).
\end{align*}

This sparse ST-DGLM naturally suggests using  \hyperlink{https://www.r-inla.org/}{\texttt{R-INLA}} \citep{rue2009approximate}, since our models fall within the class of latent Gaussian models. Although \texttt{R-INLA} is a powerful tool for modeling spatial dependence in high spatial dependence and already there are vast choices of predefined models, it has still limited support for certain temporal dependencies, as it typically relies on predefined temporal models with minimal customization options. This limitation becomes particularly pronounced in dynamic generalized linear models, where more complex structures are needed and computational demands increase substantially for certain configurations. For examples, see \cite{ruiz2012direct, ravishanker2022dynamic}. 
Instead, to approximate the resultant posterior distribution of the proposed model,  ST-DGLM (Section \ref{subsec:dens-DGLM}) and sparse ST-DGLM (Section \ref{subsec:SPDEdetails}), we opt for simulation-based MCMC inference, which theoretically provides a pseudo-exact approximation of the posterior distribution, provided the Markov chains run for a sufficient duration and fully converge. This approach offers greater flexibility, allowing customization for a wide range of models beyond those with specific response structures. In Section \ref{sec:BayesInf_prediction} below, we provide a detailed explanation of MCMC-based inference along with spatiotemporal predictions across various scenarios.

\section{Bayesian inference and spatiotemporal prediction}
\label{sec:BayesInf_prediction}
In this section, we provide the details of our proposed MCMC sampler along with a method for performing spatiotemporal predictions and inference for both the general ST--DGLM and the sparse ST--DGLM. The posterior sampling approach is largely similar for both models, with key differences in the full conditional distributions of certain parameters, particularly the space-time varying intercepts $\mu_t(\bm s)$. In this section, we provide the details of the inference procedure and spatiotemporal prediction for the sparse ST--DGLm and for the general ST--DGLM model is similar. 

\subsection{Details of MCMC}
We may express our model hierarchically (see Eqn. \eqref{eq:BYM-sparse-gen-st} in Appendix~\ref{appd:appednA}), which naturally suggests using Bayesian inference and also facilitates deriving the full conditional distributions of the model parameters (see also Section S1 in Supplementary Material). Let $\bm \Theta =  \left( \{ \bm{\lambda}_t \}_{t=1}^{T}, \{ \bm{R}_t \}_{t=1}^{T}, \{ \bm{\theta}_t \}_{t=1}^{T},\ \bm{\beta}, \tau^2,\ \sigma^2, \bm{W}, \kappa \right) $ be the model parameters (hyper + latent) of the proposed model, then the full joint posterior distribution of $\bm \Theta$ is proportional to 
  \begin{align}
  \label{eq:JointPost}
\pi(\bm \Theta \mid \bm{Y}) &\propto \prod_{t=1}^T \prod_{i=1}^n \pi(Y_t(\bm{s}_i) \mid \lambda_t(\bm{s}_i)) \times \prod_{t=1}^T \pi(\bm{\lambda}_t \mid \bm{R}_t, \bm{\theta}_t, \bm{\beta}, \tau^2) \times \prod_{t=1}^T \pi(\bm{R}_t \mid \bm{R}_{t-1}, \sigma^2, \kappa) \nonumber \\
&\quad \times \prod_{t=1}^T \pi(\bm{\theta}_t \mid \bm{\theta}_{t-1}, \bm{W}) \times \pi(\bm{\beta})\ \pi(\tau^2)\ \pi(\sigma^2)\ \prod_{l=1}^p \pi(w_l)\ \pi(\kappa). 
\end{align}
Due to the high dimensionality and complex dependencies among latent variables and hyperparameters, the resulting joint posterior does not admit a closed-form expression and is analytically intractable. While several approximate inference techniques exist in the Bayesian paradigm (e.g., variational Bayes or Laplace approximations), Markov Chain Monte Carlo (MCMC) remains the most flexible and reliable approach for posterior inference in such models. Therefore, we propose a customized hybrid MCMC sampler specifically designed for our modeling framework. This sampler is scalable and leverages model simplifications whenever possible. For instance, we assume conjugate prior distributions for model hyperparameters whenever applicable. Specifically, we assume an inverse gamma distribution for variance parameters, including \(\sigma^2\) (model variance), \(\tau^2\) (error variance), and \(W\) (state variances), leading to closed-form expressions for their full conditional distributions (see Appendix~\ref{appd:appednA} for more details). Most parameters result in closed-form full conditionals, allowing for direct Gibbs sampling, except for the log-rate parameters \(\lambda\), for which we propose using a preconditioned Metropolis-adjusted Langevin algorithm  \citep[pMALA,][]{girolami2011riemann} to define proposal distributions.  Moreover, we update the dynamic state vectors \(\theta\) in our MCMC sampler using the Forward Filtering Backward Sampling (FFBS) algorithm \citep{carter1994gibbs}, a well-established approach for sampling state vectors in dynamic linear models. We provide full details of our MCMC sampler in Appendix~\ref{appd:appednA}, along with detailed pseudocode in Algorithm~\ref{alg:mcmc-sampler-sparse}. 
 
\subsection{Spatiotemporal prediction}
Our proposed ST--DGLM and its sparse model counterpart naturally perform missing value imputations, spatial interpolation, future forecasting, and spatiotemporal predictions. Missing value imputation and spatial interpolation are performed within the observed temporal window; however, future forecasting and spatiotemporal prediction refer to predictions outside the observed temporal window. Specifically, future forecasting involves predicting at future time points at observed sites $\bm{s_i}, i=1,\ldots,n$, while spatiotemporal prediction involves predicting at unobserved sites and future time points $T+h, h=1,2,\ldots$. 

Let $\{(\bm{s}_{p_1}, t_{p_1}), \dots, (\bm{s}_{p_I}, t_{p_1}), \dots, (\bm{s}_{p_1}, t_{p_J}), \dots, (\bm{s}_{p_I}, t_{p_J})\} \in \mathcal{S} \times \mathcal{T}$ denote the unobserved segment of the spatiotemporal process at a set of arbitrary $p_I$ locations and $p_J$ time points and $
\bm Y_{\text{prd}} = \left(Y_{t_{p_1}}(\bm{s}_{p_1}), \dots, Y_{t_{p_1}}(\bm{s}_{p_I}), \dots, Y_{t_{p_J}}(\bm{s}_{p_1}), \dots, Y_{t_{p_I}}(\bm{s}_{p_J})\right)'
$ be the predictive process at these segments. Then our goal is to make conditional inferences for
$\bm Y_{\text{prd}}$ based on the observed data 
$
\bm Y = \left(Y_{1}(\bm{s}_1), \dots,Y_{1}(\bm{s}_n),\ldots,Y_{T}(\bm{s}_1),\ldots Y_{T}(\bm{s}_n)\right)'.
$ Let $\bm\xi = (\kappa, \sigma^2, \tau^2, \bm W)$ be the  hyperparameters of the model and $\bm \Psi =(\bm \mu_{\text{prd}}, \bm \mu, \bm \theta_{\text{prd}}, \bm \theta, \bm \beta, \bm \xi) \in \bm\Theta$ represent all unknown model parameters, where  $\bm \mu_{\text{prd}} = \left(\mu_{t_{p_1}}(\bm{s}_{p_1}), \dots, \mu_{t_{p_1}}(\bm{s}_{p_I}), \dots, \mu_{t_{p_J}}(\bm{s}_{p_1}), \dots, \mu_{t_{p_I}}(\bm{s}_{p_J})\right)'$, $\bm \mu = \left(\mu_{1}(\bm{s}_1), \dots,\mu_{1}(\bm{s}_n),\ldots,\mu_{T}(\bm{s}_1),\ldots \mu_{T}(\bm{s}_n)\right)'$, $\bm \theta = (\bm \theta_1', \ldots, \bm\theta_T)'$, $\bm \theta_t = (\theta_{t1}, \ldots, \theta_{tp})'$, and $\bm \theta_{\text{prd}} = (\bm \theta_{p_1}', \ldots, \bm\theta_{tp_J})'$. Then, the posterior predictive distribution is 
\begin{align}
\label{eq:pred_data_level}
\pi(\bm Y_{\text{prd}} \mid \bm Y) & = \int_{\bm \Psi} \pi_{\text{Poisson}}(\bm Y_{\text{prd}} \mid \exp(\bm \lambda_{\text{prd}})) \pi(\bm \mu_{\text{prd}}, \bm \mu, \bm \theta_{\text{prd}}, \bm \theta,   \bm \beta,  \bm \xi \mid \bm Y) \, \mathrm{d}\bm \Psi,
\end{align}
where $\bm \lambda_{\text{prd}} = \bm \mu_{\text{prd}} + \bm X_{\text{prd}}' \bm \beta + \bm F_{\text{prd}}' \bm \theta_{\text{prd}} + \bm \varepsilon$ with $\bm \lambda_{\text{prd}}$, $\bm \mu_{\text{prd}}$, $\bm X_{\text{prd}}$, and $\bm F_{\text{prd}}$ having appropriate dimensions and organized to match the space-time indices of $\bm Y_{\text{prd}}$.  The joint distribution $\pi(\bm \mu_{\text{prd}}, \bm \mu, \bm \theta_{\text{prd}}, \bm \theta, \bm \beta, \bm \xi \mid \bm Y)$ can be further simplified by factorizing the spatial and temporal components into observed and unobserved space-time points:
\begin{align}
\label{eq:pred_predictor_level} 
\pi(\bm \mu_{\text{prd}}, \bm \mu, \bm \theta_{\text{prd}} \bm \theta, \bm \beta, \bm \xi \mid \bm Y) =\pi(\bm \mu_{\text{prd}} \mid \bm \mu,  \cdot) \pi(\bm \mu \mid \cdot) \pi(\bm \theta_{\text{prd}}\mid \bm \theta, \cdot) \pi(\bm \theta \mid \cdot) \pi(\bm \beta \mid \cdot) \pi(\bm \xi \mid \cdot),
\end{align}
where `$\cdot$' in the respective conditioning distribution denotes other parameters and observed data $\bm Y$. An analytical closed-form solution is unavailable for the predictive distribution \eqref{eq:pred_data_level}. However, utilizing simulation-based Markov Chain Monte Carlo (MCMC) inference procedures, we can draw samples from the posterior predictive distribution through composition sampling. This involves generating samples of model parameters $\bm{\Psi}$ using the MCMC sampler detailed in Algorithm~\ref{alg:mcmc-sampler-sparse} in Appendix~\ref{appd:appednA}. Subsequently, for each set of  $\bm{\mu}$, $\bm{\theta}$, and related hyperparameters simulated from their respective posterior distributions, we sample from the predictive distributions $\pi(\bm{\mu}_{\text{prd}}\mid \bm{\mu}, \cdots)$ and $\pi(\bm{\theta}_{\text{prd}} \mid \bm{\theta}, \cdot)$; see Appendix~\ref{appd:appednB}. 

Now, we will briefly mention the steps to perform these predictions for a single space-time point for various prediction scenarios, namely imputations, spatial interpolation, future forecasting, and space-time prediction, which further simplifies the compositional expressions \eqref{eq:pred_data_level} and \eqref{eq:pred_predictor_level} depending on the prediction scenarios. Let $(\bm{s}_p, t_p)$ be the new space-time points where we wish to make predictions, and $T+h, h=1,2,\ldots$ be the future time points. Moreover, as previously mentioned, $\bm{s}_i, i=1,\ldots,n$, or in general $\bm{s}$, represents the spatial locations where we have observed data, and $t=1,\ldots,T$ represents the observed temporal points. 
Specifically, the prediction involves constructing the Poisson log-rate parameter $\lambda_{t_p}(\bm s_p)$, and then samples from posterior predictive distributions are obtained by sampling from the corresponding Poisson distribution with rate $\exp(\lambda_{t_p}(\bm s_p))$, i.e.,
\begin{align}
\label{eq:Post_sampling_dist_exp}
{Y}_{t_p}(\bm{s}_p) \sim \text{Poisson}\left[\exp\left\{{\lambda}_{t_p}(\bm{s}_p)\right\}\right],
\end{align}
 will be the sample from the posterior predictive distribution \eqref{eq:pred_data_level} at time $t_p$ and spatial location $\bm{s}_p$. Therefore, we may simulate $Y_{t_p}(\bm s_p)$ from the Poisson distribution for each samples of $\bm \Psi$ by first constructing $\lambda_{t_p}(\bm s_p)$ for each posterior sample of $\bm \Psi$, which then maybe used to calculate the posterior predictive mean and variances of the prediction at time $t_p$ and spatial location $\bm{s}_p$. Notably, the only difference  in performing different types of prediction are in constructing the log-rate $\lambda_{t_p}(\bm s_p)$, and the sampling from the data distribution steps \eqref{eq:Post_sampling_dist_exp} remains the same. In Appendix~\ref{appd:appednB}, we provide details on constructing the intensity $\lambda_{t_p}(\bm{s_p})$ for different prediction scenarios, namely for spatial interpolation, future forecasting, and spatiotemporal predictions. 
 \subsection{ Estimation of average annual daily bicyclist (AADB)}
\label{subsec:aadb_est} 
The Average Annual Daily Bicycle (AADB) count is a key metric in transportation studies, utilized for bike ridership analysis and infrastructure planning. At a specific site $\mathbf{s}$ and for a given year `yr', AADB is defined as:
\begin{align*}
\text{AADB}_{\text{yr}}(\mathbf{s}) = \frac{1}{n_{\text{yr}}} \sum_{t=1}^{n_{\text{yr}}} Y_t(\mathbf{s}),
\end{align*}
where $n_{\text{yr}}$ represents the number of days in a year, and $Y_t(\mathbf{s})$ denotes the bicycle count at site $\mathbf{s}$ on day $t$.
For sites with complete daily observations, calculating AADB is straightforward. However, for locations with incomplete or missing data, imputation techniques are necessary to estimate the full time series before computing AADB. Our model inherently provides a method for imputing missing values, facilitating accurate AADB calculations even with incomplete datasets.

Additionally, to estimate AADB at new sites with no observations at all, spatial interpolation methods, such as those discussed in Section \ref{subsec:kriging-sparse}, can be employed to predict the entire time series. Once the time series is estimated, calculating AADB follows the same procedure as for sites with complete data. In general, the estimated AADB at site $\mathbf{s}$ is given by:
\begin{align}
\hat{\text{AADB}}_{\text{yr}}(\mathbf{s}) = \frac{1}{n_{\text{yr}}} \sum_{t=1}^{n_{\text{yr}}} \hat{Y}_t(\mathbf{s}),
\label{eq:aadb_est}
\end{align}
where $\hat{Y}_t(\mathbf{s})$ is the estimated bicycle count at site $\mathbf{s}$ on day $t$. 
 To account for uncertainty in these estimates, $\widehat{\text{AADB}}_{\text{yr}}(\mathbf{s})$ is calculated for each simulated sample, allowing us to report point estimates and uncertainty quantification through posterior means and credible intervals for AADB based on the ensemble of simulated samples.

\section{Synthetic data experiment}
\label{sec:synthetic_exp}
In this section, we provide a synthetic data experiment to examine the trade-off between computational efficiency and accuracy of the sparse ST-DGLM proposed in Section \ref{subsec:SPDEdetails}. We also assess whether the true underlying model parameters can be recovered and evaluate the effectiveness of our MCMC sampler. The experimental data are generated using the ST--DGLM with a dense Mat\'ern covariance structure, as detailed in Section \ref{subsec:dens-DGLM}. We first simulated $n=1000$ spatial locations uniformly in $[0,1]^2$ and fixed all model hyperparameters to reasonable values (see the first row in Table \ref{tab:summTable.synexp}). We then use the model hierarchy (Eqn. \eqref{eq:BYM-sparse-gen-st} in Appendix~\ref{appd:appednA}), to simulate synthetic data for $T=200$ time points. Specifically, we use three spacetime covariates ($q=3$), simulated from a standard normal distribution, to compute the component $\bm X_t(\bm s)'\bm \beta$, along with a first-order harmonics of order 7 to calculate the component $\bm F_t'\bm\theta_t$; see Appendix~\ref{appd:appednC} for more details of these model components. Furthermore, we set the smoothness parameter to $\nu=1$ in the Mat\'ern covariance function \eqref{eq:matern}, and use the dynamics specified in \eqref{eq:gen-model-st-intcpt} to simulate the spatiotemporal intercept $\mu_t(\bm{s})$. Also, we simulate the errors $\varepsilon_{t}(\bm{s})$ from a normal distribution with mean zero and variance $\tau^2$, fixed to small values. Therefore, the log-rate parameter $\lambda_t(\bm s)$ is computed as in \eqref{eq:gen-st-model-intensity-level}, and consequently we simulated the data at spatial location $\bm s$ and time point $t$ from a Poisson distribution with rate $\exp(\lambda_t(\bm s))$. For additional details about the data simulations and model components, see Appendix~\ref{appd:appednC}.

\begin{figure}[!h]
    \centering
    \includegraphics[width=\linewidth]
    {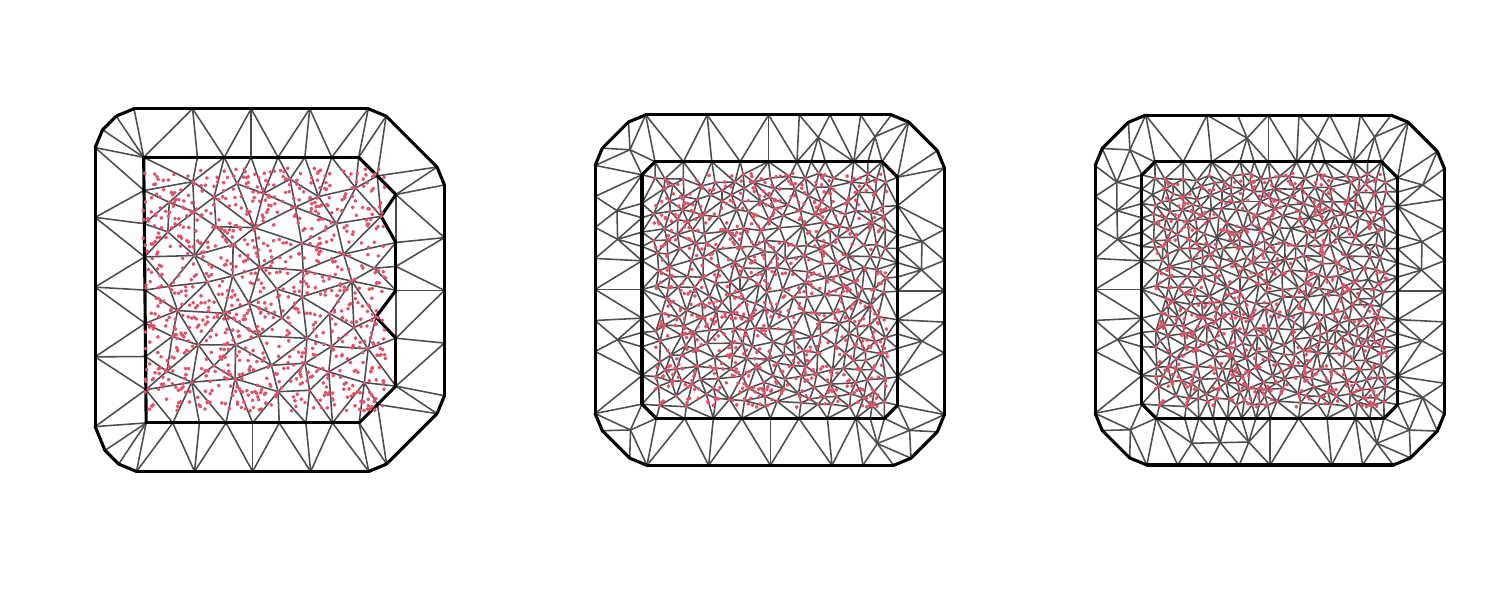}
    \vspace{-1cm}
    \caption{SPDE mesh nodes overlaid on the $n=1000$ spatial locations (red solid circles) for three choices of mesh; $N=112$ (left most), $N=376$ (middle),  $N=551$ (rightmost).  }
    \label{fig:mesh-nodes_synth_exp}
\end{figure}

We then fit ST--DGLM and its sparse counterparts detailed in 
 Sections \ref{subsec:dens-DGLM} and \ref{subsec:SPDEdetails}, respectively, to the synthetic data, using our proposed MCMC sampler in Algorithm~\ref{alg:mcmc-sampler-sparse} to generate posterior samples of the corresponding model components, including the posterior samples of $\mu_t(\bm s)$ for the ST--DGLM and $R_t(\bm s^\star)$ for its sparse counterpart, which is then used to obtain the approximate intercepts $\mu_t(\bm{s}) = \bm{a}(\bm{s})' \bm{R}_t = \sum_{k=1}^{N} a(\bm{s}, \bm{s}_k^*) R_{t}(\bm s_k^*)$. We generated 20,000 MCMC samples, discarded 15,000 as burn-in, and thinned the remaining 5,000 by a factor of 10, resulting in 500 samples for the posterior summary. To evaluate the convergence of Markov chains, we ran four parallel Markov chains with different initial values.  Visual inspections of the trace plots suggest that the chains quickly converge to their stationary distributions and show good mixing. We fit the sparse ST--DGLM using various mesh configurations to assess the impact of mesh choice on model performance.  Panels of Figure \ref{fig:mesh-nodes_synth_exp} show the mesh locations around the study region for three choices of mesh node configuration, ranging from fewer mesh nodes (less accurate but computationally cheap) to more mesh nodes (more accurate and relatively computationally costly). For both the ST-DGLM and its sparse model counterparts, we utilized a single node with one CPU core on the Narval cluster, which is part of the Digital Research Alliance of Canada.

Table \ref{tab:comparison_synth_exp} provides a summary of the estimated intercept parameters for the ST–DGLM and its sparse counterparts for three different choices of mesh. We notice that as the number of mesh nodes increases, the accuracy of the intercept parameter estimates also improves, thus we gain in computational efficiency but lose in efficiency. However, this comes at the cost of higher computational expenses, highlighting the need to strike a balance between computational efficiency and estimation accuracy. Notably, it is well-established that moderate mesh sizes combined with well-designed mesh constructions generally yield valid improvements in accuracy and the choice of mesh often has a limited impact on the results. For detailed discussions on mesh construction techniques, see Chapter~2 of \cite{krainski2018advanced}. 

\begin{table}[!h]
\centering
\renewcommand{\arraystretch}{1.1}
\caption{Root mean squared error (rMSE), mean absolute error (MAE), continuous ranked probability score (CRPS), and relative computational gain (compared to the ST--DGLM)  for the competing models in the synthetic data experiment.}
\label{tab:comparison_synth_exp}
\begin{tabular}{@{}lrrrr@{}}
\toprule
Model & rMSPE & MAE & CRPS & Relative Gain \\ \midrule
ST--DGLM          & 0.294 & 0.228 & 0.172   & -- \\
Sparse ST--DGLM $N=551$ & 0.291 & 0.225 & 0.169  & 5.434 \\
Sparse ST--DGLM $N=376$ & 0.308 & 0.243 & 0.184  & 11.230 \\
Sparse ST--DGLM $N=112$ & 0.312 & 0.242 & 0.177  & 12.396 \\ \bottomrule
\end{tabular}
\end{table}

\begin{figure}[t!]
    \centering
    \includegraphics[width=0.8\linewidth]{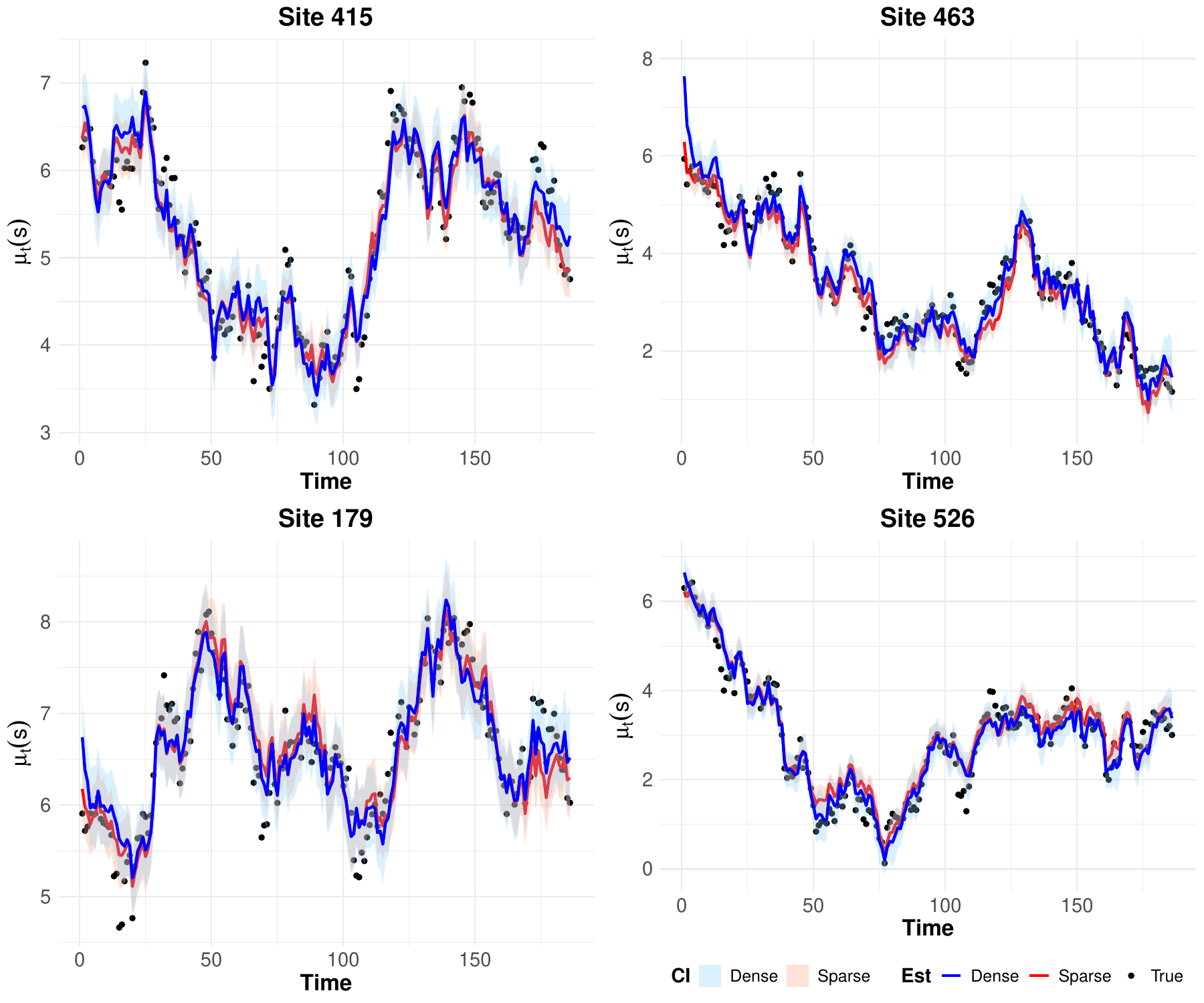}
     \caption{Estimated space-time varying intercepts along with 95\% credible intervals for ST--DGLM with dense covariance structure and its sparse model counterpart in synthetic data experiment. Here black dots denote the true intercept values.   }
    \label{fig:synth_exp_intcp_compare}
\end{figure}

Figure \ref{fig:synth_exp_intcp_compare} shows the time series of true intercept values alongside the estimated intercepts and credible intervals for the ST--DGLM model and its sparse counterparts with a mesh size of $N = 360$, for four randomly selected sites. From this, it is evident that the true intercept values are well-recovered, further validating both the efficacy of our MCMC sampler (Algorithm \ref{alg:mcmc-sampler-sparse}) and the accuracy of the sparse ST--DGLM model in estimating the true intercept parameters.
 This claim is further supported by the spatial interpolation plots in Figure~S1 of the Supplementary Material (SM), where we randomly removed 10 sites from the observed set of locations and performed spatial interpolation for these locations. The interpolated counts for both approaches are nearly indistinguishable in terms of their estimates and posterior predictive credible intervals. A similar pattern is observed in the forecasting plots and spatiotemporal prediction plots shown in Figures S2 and S3 of the Supplementary Materials, respectively. Moreover, we artificially introduced a missing pattern where observations were randomly removed from some sites, incorporating random missing block structures. From Figure~S4 of the Supplementary Material, it is evident that these missing patterns are well recovered, thereby validating the effectiveness of our MCMC sampler and the performance of our model in missing value imputation.  We also wanted to assess whether we are able to recover the time-evolving coefficients corresponding to purely temporal covariates. Figure~S5 in the Supplementary Material shows the estimated coefficients along with the true values and their 95\% credible intervals which confirms that we are able to recover the true pattern. 
 
 Table~\ref{tab:summTable.synexp} presents the posterior means of all model hyperparameters and fixed covariate coefficients alongside their corresponding true values, standard deviations of the estimates, and associated 95\% credible intervals. 
In general, for most of the coefficients and hyperparameters, the true values used to generate the data are successfully recovered. Most parameter estimates are accurate, except the variance hyperparameters ($\sigma^2$ and $\tau^2$), which are notably harder to estimate. Nevertheless, we achieve satisfactory estimates for other latent model components, such as space-time varying intercepts, as shown in Figure \ref{fig:synth_exp_intcp_compare}, and coefficients of purely temporally varying components (Figure~S5 in the Supplementary Material). This indicates that a slight mismatch in the estimation of some parameters does not significantly affect the estimation of the latent components.

\begin{table}[!h]
\centering
\renewcommand{\arraystretch}{1.1}
\caption{Posterior summary of model hyperparameters and covariate coefficients in synthetic data experiments, including the posterior mean, 95\% credible interval (CI), and true values.}
\label{tab:summTable.synexp}
\begin{tabular}{lccc}
\toprule
Parameter & True value & Mean & 95\% CI \\
\midrule
$\kappa$    & 0.350 & 0.356 & \footnotesize{[0.342, 0.372]} \\
$\sigma^2$  & 0.100 & 0.058 & \footnotesize{[0.056, 0.061]} \\
$\tau^2$    & 0.050 & 0.070 & \footnotesize{[0.069, 0.070]} \\
$w_1$       & 0.010 & 0.023 & \footnotesize{[0.012, 0.038]} \\
$w_2$       & 0.020 & 0.095 & \footnotesize{[0.052, 0.142]} \\
$\beta_1$   & 0.266 & 0.266 & \footnotesize{[0.265, 0.266]} \\
$\beta_2$   & 0.372 & 0.372 & \footnotesize{[0.371, 0.373]} \\
$\beta_3$   & 0.573 & 0.573 & \footnotesize{[0.573, 0.573]} \\
\bottomrule
\end{tabular}
\end{table}

Overall this synthetic data experiment shows that the sparse ST-DGLM model provides good computational efficiency while maintaining accuracy very similar to the ST--DGLM model with dense covariance structures. 

\section{Application to Eco-counter bicycle  counts data}
\label{sec:application}
\subsection{Data description and exploratory analysis}
\label{subsec:EDA}
The Eco-counter bicycle dataset includes hourly bicycle count data observed at fixed installed sensors within the Montreal Island region. This data includes 53 long-term sites (censors) across Montreal Island, spanning from January 1, 2017, to August 31, 2023. In addition to bicycle count data, the dataset includes the coordinates of each site, along with supplementary site-specific information, such as an indicator specifying whether a site is located at an intersection or along a link. Some stations exhibit partially incomplete temporal data, with most of the missingness occurring at the beginning of the observation period due to faulty or inactive sensors. Weather-related variables such as temperature, humidity, rainfall, and wind speed at each site were obtained from the \href{https://xn--montral-fya.weatherstats.ca/}{Montr\'eal Airport} weather station at an hourly scale for the same period as the bicycle  count data. Given Montr\'eal's relatively small geographic area, these weather variables do not vary significantly across sites, allowing us to treat them as purely temporal variables. Due to Montr\'eal's winter conditions, the bicycle count data from mid-November to mid-April contains many zeros. Therefore, we focused on data from April 15 to November 15 each year, known as the ``bicycle season" when most bike companies such as BIXI \citep{bixi2023} operate. Additionally, we aggregated the temporal resolution of the data from hourly to daily counts, summed over 24 hours. Finally, for our analysis, we use data from 46 sites where at least 90\% of the data was available for the years 2021 and 2022, to demonstrate the applicability of our modeling framework.

The exploratory analysis (see plots in Section  S3 of the Supplementary Material) suggests weekly periodic cycles and a strong autocorrelation at lag one. Seasonal cycles are also clear, reflecting the varying use of bikes across different months each year, and there also seem to be slight weekday and weekend effects. The weather variables, such as temperature, precipitation, wind speed, relative humidity, and visibility, appear to be mostly uncorrelated, indicating no issues of multicollinearity, which is favorable as it allows us to include these variables in the design matrix $\bm X_t(\bm s)$ to capture their effects. 

\subsection{Competing models}
\label{subsec:competing_models}
% \subsubsection{Our proposed ST-DGLM and sparse ST--DGLM }
The exploratory analysis in Section \ref{subsec:EDA} provides insights into potential model components that are well-suited for our application. In particular, models with components designed to capture weekly seasonality and accommodate autoregressive patterns, such as random walks, emerge as natural choices, and the general modeling framework detailed in Section \ref{subsec:dens-DGLM} is particularly well suited to include such features. Additionally, we incorporate harmonics of order seven with purely temporally varying coefficients to capture weekly seasonality. We include purely temporal variables such as temperature, visibility, wind speed, and precipitation as explanatory or predictor variables in the fixed-effects $\bm X_t(\bm s)$. We also include purely spatial variables, such as elevation, walk scores, and population density, in $\bm X_t(\bm s)$, along with these variables, w. In the Mat\'ern \eqref{eq:matern} we set the smoothness parameters fixed to $0.5$, which results in an exponential covariance function. A similar set of model components is also adopted for the sparse ST--DGLM.  For a detailed description of these modeling frameworks and explanations of individual model components, see Appendix~\ref{appd:appednD}. The results for ST--DGLM, and sparse ST--DGLM were obtained using the Narval cluster, part of the Digital Research Alliance of Canada. Computations were performed on a single node with one CPU core for all these models. We ran both models for 20,000 iterations, using the first 15,000 as burn-in. After thinning by a factor of 10, this resulted in a final posterior sample of 500 for summary and inference. Generating these 20,000 samples took approximately 1.5 hours. Additionally, we ran four independent Markov chains to assess convergence, which appeared to be achieved quickly, with satisfactory mixing, see Figures S16–S20 in the Supplementary Material.

In our analysis, we include a Bayesian generalized linear model (GLM) with a Poisson response as a baseline model. In this model,  count data are assumed to   follow a Poisson distribution with log-rate parameter, $\lambda_t(\bm s)$ such that:  $
\lambda_t(\bm s) = \mu + \bm X_t(\bm s)'\bm \beta + \varepsilon_t(\bm s)
$, where $\varepsilon_t(\bm s)$ follows a normal distribution with mean zero and variance $\tau^2$.  In the design matrix $\bm X_t(\bm s)$, we include all the predictor variables as in  ST--DGLM and sparse ST--DGLM. This simplified model is easy to implement and provides a useful reference point for evaluating the performance of more complex competing models.

Another competing model we incorporate is the BKTR model proposed by \cite{lei2024scalable} and implemented in the \texttt{R} and \texttt{Python} packages described in \cite{lanthierbktr}. BKTR is an efficient spatiotemporal regression implementing a tensor regression approach,
which greatly reduces the model's computational cost. The framework also assumes
Gaussian process (GP) priors to capture the spatial and temporal dependencies of the data in a Bayesian context. In our analysis, we selected all the covariates in the design matrix \(\bm{X}_t(\bm{s})\) that were also used in the ST-DGLM and its sparse model counterparts, and set the tensor rank to 15. The bicycle count data was log-transformed to better meet with the assumptions of normal response of the BKTR. The algorithm was run for  2000 MCMC iterations, with a burn-in period of 1500 iterations. We also tested the BKTR model with different combinations of tensor rank and MCMC iterations, but the results remained consistent. The BKTR model was run on a personal computer with the following configuration: macOS Sonoma 14.5, Apple M3 chip with an 8-core CPU, integrated GPU, and 8 GB of RAM. 

The last model included in our analysis is the Bayesian Neural Field \citep[BayesNF,][]{saad2024scalable} as a competing model, which is a scalable spatiotemporal modeling framework that integrates deep neural networks with hierarchical Bayesian modeling, making it highly efficient for handling large datasets. In this model, observations $ Y_t(\bm{s})$ are modeled using a probabilistic observation model:
$
Y_t(\bm{s}) \sim \text{Dist}\left\{g(F_t(\bm{s})), \gamma\right\},
$ where $g$ is a link function, $\text{Dist}$ represents distributions such as Gaussian, and $\gamma$ are the hyperparameters. The latent process $F_t(\bm{s})$ is expressed as:
$
F_t(\bm{s}) = h(\bm X_t(\bm{s}); \beta) + \eta_t(\bm{s}),
$
where $h(\bm X_t(\bm{s}); \bm \beta)$ defines the systematic mean captured by the neural network architecture. This architecture processes inputs driven by covariates $ \bm X_t(\bm{s}) $ and its several components, with coefficients $ \bm \beta $. The term $\eta_t(\bm{s})$ models the residual spatiotemporal dependencies using a Gaussian process, allowing the model to capture variability not explained by covariates. Posterior inference is performed using variational inference and maximum a {\it posteriori} estimation, both of which allow the model to scale linearly with the size of the dataset. This enables BayesNF to efficiently handle missing data, quantify uncertainty, and capture flexible dependencies. BayesNF is implemented as an open-source software package, available at:
$
\texttt{https://github.com/google/bayesnf}.
$
Finally, it supports high-performance computation using GPUs and TPUs, making it a powerful framework for large-scale spatiotemporal applications. In our analysis, we utilized all available covariates in the design matrix \(\bm{X}_t(\bm s)\) and included two harmonics to account for weekly seasonality, as implemented in the ST-DGLM and its sparse model counterparts. For the observation model, we selected the normal distribution applied to the log-transformed bicycle counts. Results were generated using a Google v2-8 TPU, requiring approximately five minutes to compute. The layer width was set to 512, the depth to 2 (both default settings), the ensemble size to 64, and the algorithm was run for 10,000 epochs.

\subsection{Results}
\label{subsec:results}
We fit all the competing models detailed in Section \ref{subsec:competing_models}, along with our proposed ST-DGLM and its sparse model counterpart to the Eco-counter dataset. We removed the data from four sites entirely assess the prediction at unobserved spatial locations. Additionally, we use varying proportions of missing values, 50\%, 60\%, and 80\%; at four other sites to evaluate how the extent of missing data affects predictions at observed locations; see Figure~S9 in Supplementary Material for the location of these sites. We use summary measures such as mean absolute error (MAE), root mean squared error (rMSE), and continuous ranked probability score \citep[CRPS,][]{gneiting2007strictly} to evaluate prediction performance.
\begin{table}[t!]
\centering
\caption{
Summary of prediction performances, including missing values imputation and spatial interpolation, for all competing models fitted to the Eco-counter dataset under three different missing data proportions (50\%, 60\%, and 80\%). The bold highlighted numbers are the best (minimum) for each missing proportion.} 
\label{tab:appl_competing_models}
%\scriptsize
\renewcommand{\arraystretch}{1.1} 
\begin{tabular}{llccc}
  \hline
Model Type & Missing Percentage & MAE & rMSE & CRPS \\
  \hline
\multirow{3}{*}{Poisson GLM} & 50 & 433.41 & 564.12 & 859.26 \\
                            & 60 & 442.18 & 574.14 & 865.48 \\
                            & 80 & 445.86 & 579.17 & 892.55 \\
\hline
\multirow{3}{*}{ST--DGLM}     & 50 & \textbf{268.25} & \textbf{341.69} & \textbf{180.793} \\
                            & 60 & 339.49 & 450.79 & \textbf{235.68} \\
                            & 80 & 361.21 & 496.07 & \textbf{252.411} \\
\hline
\multirow{3}{*}{sparse ST--DGLM}    & 50 & 280.41 & 357.50 & 335.71 \\
                            & 60 & \textbf{309.82} & \textbf{393.26} & 341.039 \\
                            & 80 & \textbf{347.94} & \textbf{459.30} & 353.65 \\
\hline
\multirow{3}{*}{BayesNF}    & 50 &  570.32 &  570.32 & 411.90 \\
                            & 60 & 662.79 &799.71 & 458.29 \\
                            & 80 & 671.72 & 801.92 & 449.61 \\
\hline
\multirow{3}{*}{BKTR}    & 50 &  396.53 &  511.16 & -- \\
                            & 60 & 410.06 & 541.67 & -- \\
                            & 80 & 438.55 & 614.25 & -- \\
\hline
\end{tabular}
%\end{adjustbox}
\end{table}
\begin{figure}[t!]
    \centering
    \includegraphics[width=0.8\linewidth]{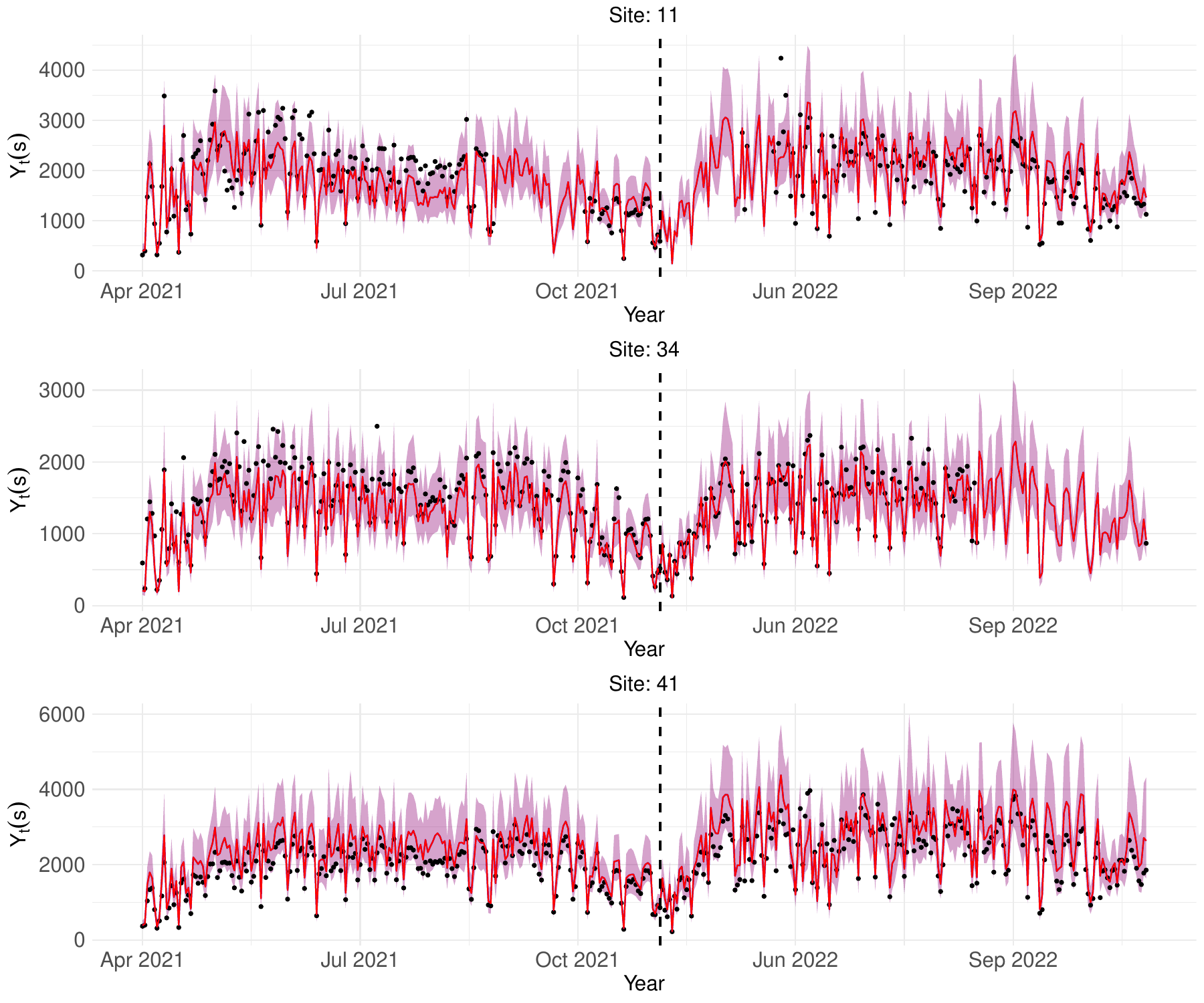}
    \caption{Spatial interpolation plots for three sites with completely removed data, showing true values (black solid circles) and 95\% credible intervals based on the best-performing model, dense ST-DGLM, under 50\% missingness. }
    \label{fig:spatIntpl_Eco}
\end{figure}
 Table \ref{tab:appl_competing_models} provides the summary metrics for all competing models across different missing value proportions. As expected, for all models, prediction errors (MAE, rMSE, and CRPS) increased as the proportion of missing data increased. The general ST--DGLM consistently achieved the lowest prediction errors, with the minimum MAE and rMSE values at 50\% missingness. Furthermore, it has the lowest CRPS scores among all models, making it the most effective choice for missing value imputation and spatial interpolation, two critical components for calculating the AADB at sites with missing data and the sites requiring spatial predictions. The sparse ST--DGLM model performed comparably to its dense counterpart, showing slightly smaller MAE and rMSE values at 60\% and 80\% missingness. However, it has slightly higher CRPS scores due to increased prediction uncertainty. Nevertheless, the sparse ST--DGLM remains reliable and is the second-best performing model among all competing models. The third-best model is the BKTR model of \citet{lei2024scalable}, which achieved competitive MAE and rMSE values. However, since BKTR does not currently support the estimation of prediction standard deviations or credible intervals, we could not compute its CRPS scores. The fourth-best model was the BayesNF model of \citet{saad2024scalable}. Despite larger MAE and rMSE values, it has relatively lower CRPS scores compared to the baseline Poisson GLM. This improvement in CRPS can be attributed to narrower prediction intervals in the BayesNF model. 

\begin{figure}[t!]
    \centering
    \includegraphics[width=0.8\linewidth]{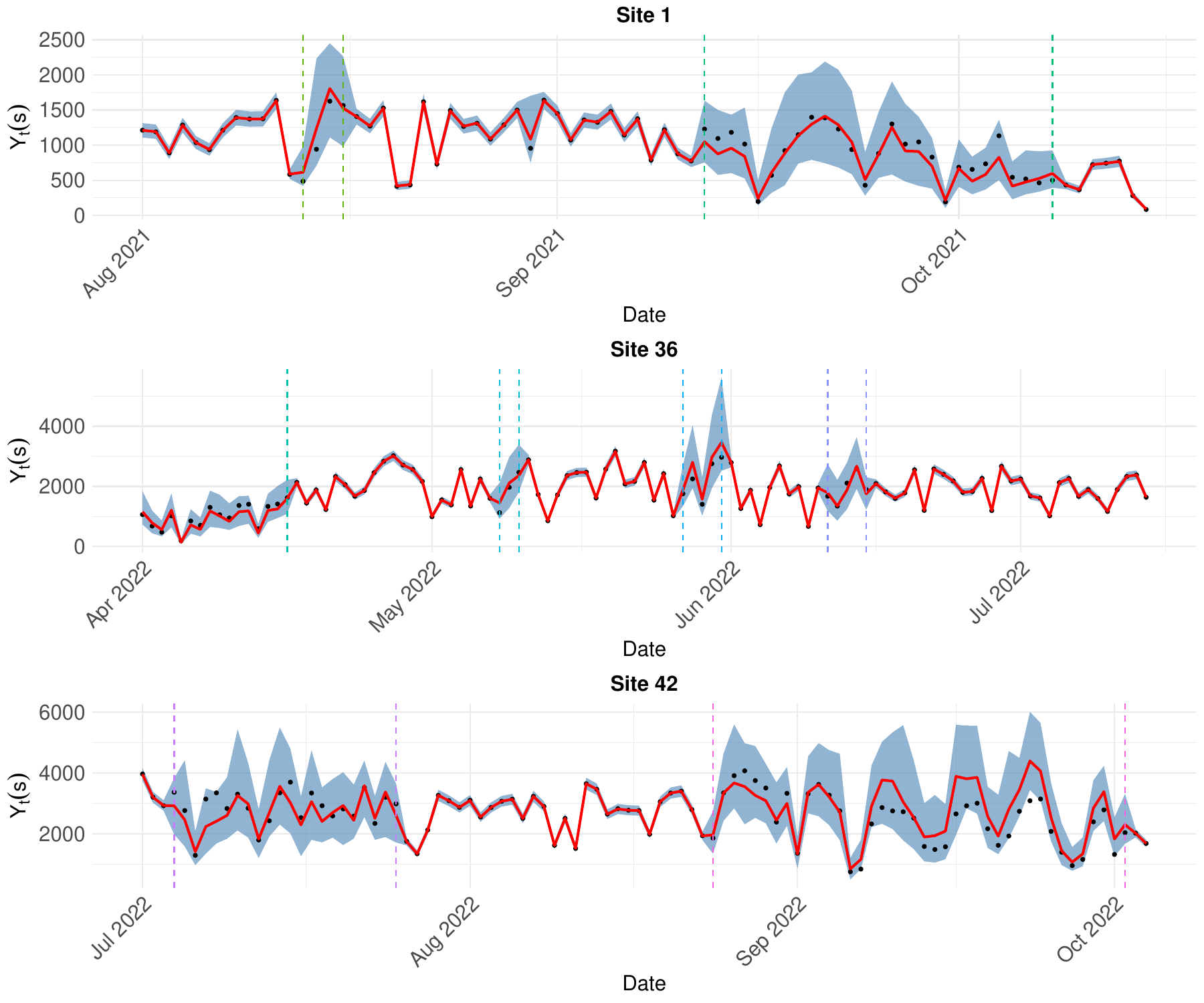}
    \caption{Missing value imputations for three sites where 50\% of the data is removed randomly in selected blocks of varying length and time periods. The figure displays true values (black solid circles) alongside 95\% credible intervals based on the ST--DGLM. Each pair of vertical dashed lines in the same color represents a time block where the time series data is completely missing. For better visibility of time structures, only parts of time series are shown.  }
    \label{fig:miss_estimation_Eco}
\end{figure}
Figure \ref{fig:spatIntpl_Eco} provides the summaries (mean and limits of the 95\% credible intervals) of the predictive posterior distributions  for the time series at  sites that had the observations removed to check predictive performance of the best model (ST--DGLM) among the fitted ones. 
The results show strong predictive performance at unobserved spatial locations, with relatively narrow credible intervals that effectively capture the underlying structure of the time series. This highlights the capability of our modeling framework to accurately replicate spatial and temporal patterns, further validating its effectiveness in handling prediction tasks at unknown spatial locations. 

Figure \ref{fig:miss_estimation_Eco} illustrates the estimation of missing values at sites where 50\% of the data were removed in random blocks of varying lengths. From this plot, it is evident that the imputation of missing values is well-recovered, and the dynamics of the time series are accurately captured.

\begin{figure}[t!]
    \centering
    \includegraphics[width=1\linewidth]{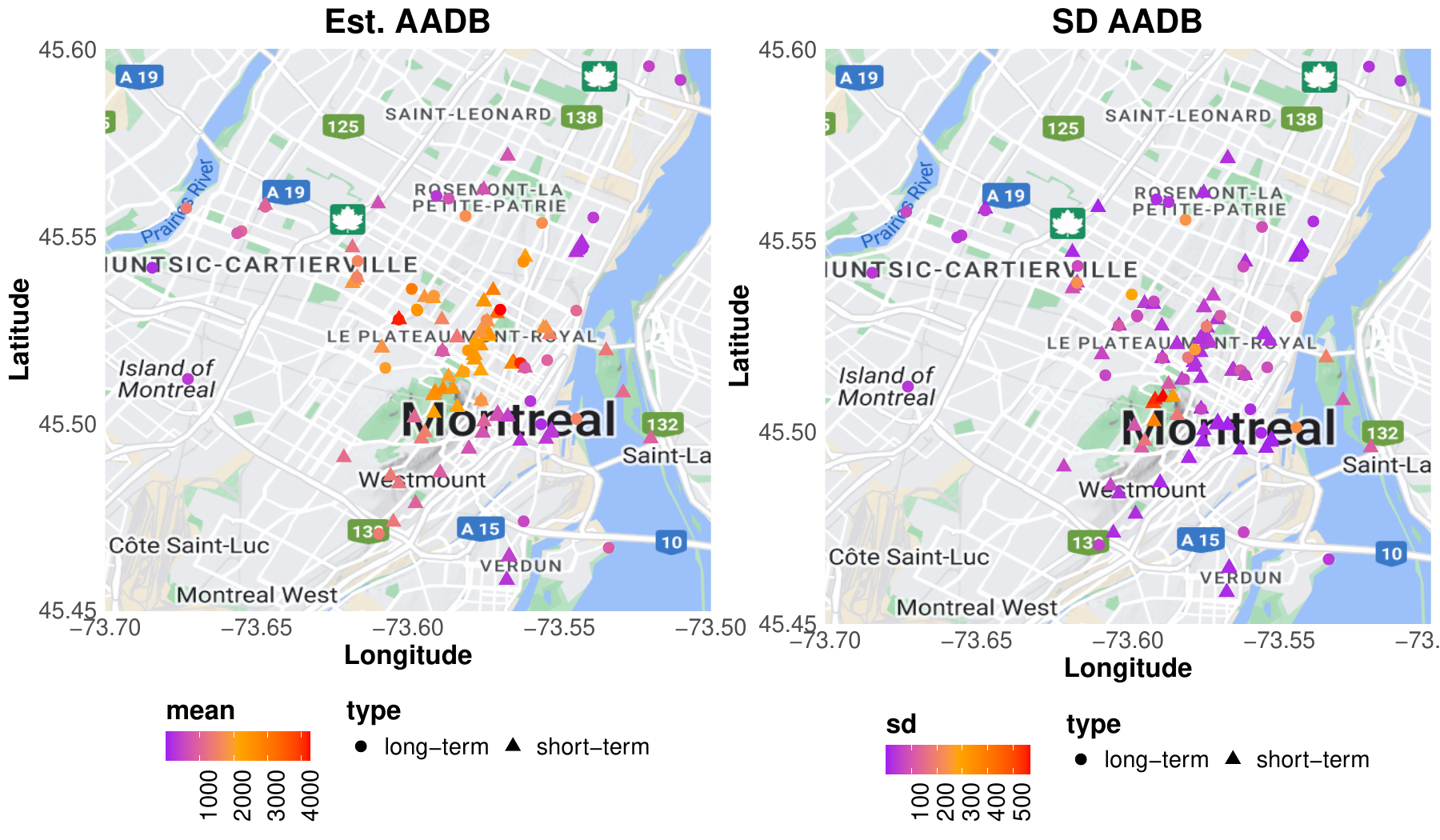}
   
    \caption{Estimated AADB (left), standard errors (right) for both long-term and short-term sites based on the best model, the ST--DGLM. }
    \label{fig:aadbEst_Eco}
\end{figure}

One of the key quantities of interest in our study is the estimation of AADB at sites where data were observed partially or removed completely. In addition to the 53 long-term monitoring sites, we also have partial data from another 93 sites, where records were collected for very short durations, sometimes as little as one or two days to provide an initial understanding of bicycle demand at those locations. These sites exhibit extremely high proportions of missing values, often exceeding 99\%. Therefore, we consider these sites as completely unobserved in our study and perform spatial interpolations based on the best-fitted model, the ST--DGLM. Using the imputed data and spatially interpolated data, we obtain the estimate of AADB and standard errors for each site based on posterior samples, following the expression in \eqref{eq:aadb_est}. Figure \ref{fig:aadbEst_Eco} shows the estimated AADB along with their standard errors and credible intervals for both types of sites; long-term and short-term. The results seems consistent with the empirical analysis, with higher bicycle demand observed near Mont Royal park and major universities, likely reflecting the high density of students and recreational users in these areas. In contrast, sites located on the outskirts of Montreal exhibit significantly lower bicycle counts. This pattern is also evident at the short-term monitoring sites, aligning with initial expectations, as these stations are intuitively located in areas with less frequent bicycle use. 

\begin{table}[t!]
\centering
\renewcommand{\arraystretch}{1.1}
\caption{
Posterior summary of model hyperparameters and covariate coefficients for the ST--DGLM, including posterior means, medians, and 95\% credible intervals (CI).
}
\begin{tabular}{lccc}
  \hline
\multicolumn{4}{c}{Posterior summary of Matérn and error variances hyperparameters} \\
  \hline
Parameter & Mean & Median & 95\% CI \\
  \hline
$\kappa$ (spat. range) & 5.0845 & 5.0715 & [4.6340, 5.6548] \\
$\sigma^2$ (var. intcpt.) & 0.0842 & 0.0839 & [0.0772, 0.0911] \\
$\tau^2$ (var. nugget) & 0.0159 & 0.0159 & [0.0148, 0.0169] \\
  \hline
\multicolumn{4}{c}{Posterior summary of variances of state-space parameters} \\
  \hline
Parameter & Mean & Median & 95\% CI \\
  \hline
$w_1$ & 0.0133 & 0.0128 & [0.0073, 0.0227] \\
$w_2$ & 0.0119 & 0.0114 & [0.0074, 0.0189] \\
$w_3$ & 0.0188 & 0.0183 & [0.0112, 0.0306] \\
$w_4$ & 0.0292 & 0.0291 & [0.0181, 0.0424] \\
  \hline
\multicolumn{4}{c}{Posterior summary of covariate coefficients} \\
  \hline
Covariate & Mean & Median & 95\% CI \\
  \hline
$\beta_{T_t}$ & 0.2863 & 0.2905 & [0.2452, 0.3051] \\
$\beta_{v_t}$ & 0.2996 & 0.3003 & [0.2752, 0.3184] \\
$\beta_{w_t}$ & -0.0354 & -0.0335 & [-0.0639, -0.0144] \\
$\beta_{pd_t}$ & -0.3610 & -0.3603 & [-0.3978, -0.3253] \\
$\beta_{wd_t}$ & 0.0266 & 0.0272 & [-0.0323, 0.0661] \\
$\beta_{yr_t}$ & -0.1347 & -0.1413 & [-0.1865, -0.0749] \\
$\beta_{elv(\bm s)}$ & 0.0969 & 0.0886 & [0.0267, 0.1929] \\
$\beta_{ws(\bm s)}$ & -0.1312 & -0.1256 & [-0.1784, -0.0894] \\
$\beta_{pp(\bm s)}$ & -0.1271 & -0.1262 & [-0.1618, -0.0842] \\
  \hline
\end{tabular}
% \end{adjustbox}
% \endgroup
\label{tab:summ_param_est}
\end{table}

Table \ref{tab:summ_param_est} provides the posterior summary of all hyperparameters and covariate coefficients estimated using the ST--DGLM. The table includes posterior means, medians, standard deviations, and 95\% credible intervals for the model parameters. 
The estimated Mat\'ern range parameter is approximately $5\text{km}$, indicating moderate spatial dependence. The variance of the process is $\hat{\sigma^2} = 0.084$, suggesting moderate marginal variations in the intercept terms. As expected, the variance of the independent random components $\varepsilon$ is estimated to be very small $(\hat{\tau^2} = 0.015)$, further demonstrating that the model components effectively capture the spatiotemporal patterns, leaving minimal residual variation or unexplained noise. Moreover, since the estimated $\hat{\sigma}^2 > \hat{\tau}^2$, this suggests that the structured spatiotemporal process accounts for substantially more variation than the unstructured (white noise) component. 
The estimated variance of the temporally varying coefficients are also very small, suggesting a smooth evolution of $\theta$'s. Consistent with exploratory analysis and intuitive reasoning, temperature and visibility have strong positive effects on bicycle usage, with estimated coefficients of \(\hat{\beta}_{T_t} = 0.28\) and \(\hat{\beta}_{v_t} = 0.30\), respectively. In contrast, precipitation (modeled as a dummy variable) has a significantly negative effect, with an estimated coefficient of \(\hat{\beta}_{pd_t} = -0.38\). Other significant purely spatial covariates include walk scores (\(\hat{\beta}_{walk} = -0.13\)) and population density (\(\hat{\beta}_{pp} = -0.13\)), both exhibiting negative effects. These findings suggest that higher walk scores correlate with reduced bicycle usage, likely due to better pedestrian infrastructure, while denser neighborhoods exhibit lower bicycle activity, potentially due to alternative transportation preferences or space constraints.

\section{Summary and future research direction}
\label{sec:conclusion}
We propose a spatiotemporal dynamic generalized linear modeling (ST-DGLM) framework for spatiotemporal prediction and inference of bicycle counts. To address computational challenges in large spatiotemporal datasets, we extend this framework using sparse precision matrices based on the SPDE approach of \cite{lindgren2011explicit}. With the sparse ST--DGLM framework, we achieve significant computational speedups and accuracies are competitive from its dense counter-parts. Although, in this study, our primary focus is on Eco-counter dataset with relatively small dimensions, the sparse framework is highly generalizable, making it suitable for larger datasets while offering notable computational advantages.

For the inference procedure, we develop a customized MCMC sampler that integrates multiple existing MCMC techniques into a hybrid sampler. This tailored approach enables to achieve fast and scalable performance for moderate to high spatiotemporal dimensions. Validation with synthetic data applications highlights the framework’s ability to deliver accurate predictions and substantial computational gains within reasonable time limits. Through the comparative study against recent \texttt{R} package BKTR, Python package BayesNF, as well as the baseline Poisson GLM model, we demonstrate that the ST-DGLM framework consistently outperforms alternatives, with the sparse ST-DGLM models achieving competitive results. 

A notable strength of our framework is its ability to achieve strong predictive performance while maintaining the interpretability of its components. The unified Bayesian approach facilitates the simultaneous modeling of time trends, seasonal effects, and covariate impacts, while  providing straightforward uncertainty quantification. Moreover, our framework directly supports missing value imputation, spatial interpolation, forecasting, spatiotemporal prediction, along with the predictions of AADB at short-term counting sites or unknown locations, further enhancing the applicability of our modeling framework. In this work, we apply our modeling framework to Eco-counter bicycle data, however, it is broadly applicable to other types of transportation data, including datasets with varying periodicity, temporal trends, and covariate effects. In addition, our modeling framework is implemented in an \texttt{R} package \texttt{sparseDGLM}, making it accessible and adaptable for use with various transportation datasets that share similar structures, such as time trends, seasonality, and covariate relationships. 

Our proposed framework is general and effective for inference and spatiotemporal prediction of bicycle counts. However, it has certain limitations. One significant avenue for future work is to adapt the framework to better incorporate road structures and connections between monitoring sites \citep{borovitskiy2021matern}. Such an implementation would enhance the applicability and efficiency of the proposed modeling approach and potentially improve both prediction accuracy and interpretability.  

\section*{Acknowledgments}  
 We extend our gratitude to Eduardo Adame Valenzuela for providing the initial dataset and to David Beitel for his valuable contributions through periodic discussions. We also thank EcoCounter, a company specializing in automated pedestrian and cyclist counting solutions, for providing access to the data. Additionally, we acknowledge the Digital Research Alliance of Canada and the Narval cluster for providing computational resources.

\section*{Data and Code Availability}  
The development version of our \texttt{R} packages \texttt{sparseDGLM} used in this study is currently available locally and can be accessed upon request. We plan to host the packages on GitHub once the paper reaches its final stages. Additionally, we aim to include a detailed vignette with a step-by-step guide for fitting these models, which will be made public alongside the paper as the review process nears completion.

\section*{Supplementary Material}  
For supplementary information of this article, please see \href{https://drive.google.com/file/d/1d4YBoIgOYDQfywbqSNOVhHNfQ2mEkyuS/view}{Supplementary Material}.

\section*{Funding}  
This work was supported by the CANSSI Collaborative Research Team (CRT) grant and the IVADO-FRQ Research Chair in Data Science. Partial funding was also provided by NSERC.  Schmidt is grateful for financial support from the Natural Sciences and Engineering Research Council (NSERC) of Canada (Discovery Grant RGPIN-2024-04312).
\section*{Declarations}  
\textbf{Conflicts of Interest:} The authors declare no conflicts of interest.

\baselineskip 14pt
\newpage
\bibliographystyle{CUP}
\bibliography{ref}
\newpage
\appendix
\renewcommand{\theequation}{\Alph{section}\arabic{equation}}
\section{Details of MCMC sampler}
\setcounter{equation}{0}
\label{appd:appednA}
Assume that the bicycle counts are observed at the $n$ spatial locations $\bm{s}_1, \ldots, \bm{s}_n$ and time points $t = 1, \ldots, T$. Representing Bayesian spatiotemporal models using vectorized notation often simplifies the derivation of full conditionals for parameters, especially when handling multivariate Gaussian vectors in posterior calculations. 
Let $\bm{Y}_t = (Y_t(\bm{s}_1), \ldots, Y_t(\bm{s}_n))'$ be the random vector at time $t = 1, \ldots, T$ across $n$ locations, each marginally following a Poisson distribution with corresponding rate vector $\bm{\lambda}_t = (\lambda_t(\bm{s}_1), \ldots, \lambda_t(\bm{s}_n))'$. Also, let the matrix $\bm{A} = [\bm{a}(\bm{s}_1)^{'}, \ldots, \bm{a}(\bm{s}_n)^{'}]$ be formed by appending the vectors $\bm{a}(\bm{s}_i), i = 1, \ldots, n$ row-wise. The matrix $\bm{A}$ is therefore an $n \times N$ projection matrix that projects the spatial random field $\bm{R}_t$ to the observed locations, resulting in $\bm{\mu}_t = \bm{A} \bm{R}_t$, where $\bm{\mu}_t = (\mu_t(\bm{s}_1), \ldots, \mu_t(\bm{s}_n))^{'}$. The projection matrix $\bm{A}$ is implemented in the function \texttt{inla.spde.make.A} from the \hyperlink{https://www.r-inla.org/}{\texttt{R-INLA}} package. Using these notations, the sparse ST--DGLM described in Section \ref{subsec:SPDEdetails} may be written hierarchically as: 
\begin{align}
\label{eq:BYM-sparse-gen-st}
Y_t(\bm{s}_i) \mid \bm{\lambda}_t &\stackrel{\text{ind}}{\sim} \text{Poisson}\left[\exp\left\{\lambda_t(\bm{s}_i)\right\}\right],\quad i=1,\ldots,n, t=1,\ldots,T; \\
\bm{\lambda}_t \mid \bm{R}_t, \bm{\beta}, \tau^2, \bm{\theta}_t &=  \mathcal{N}_n\left\{\bm{A} \bm{R}_t +  (\bm{F}_t' \bm{\theta}_t) \bm{1}_n + \bm{X}_t \bm{\beta},\quad \tau^2 \mathbb{I}_n \right\}; \nonumber \\
\bm{R}_t \mid \bm{R}_{t-1}, \sigma^2, \kappa &\sim \mathcal{N}_N\left\{\bm{R}_{t-1}, \sigma^2\bm{Q}_\kappa^{-1} \right\}, \quad t=1,\ldots,T, \quad \bm{R}_0 = \bm 0; \nonumber \\
\bm{\theta}_t \mid \bm{\theta}_{t-1}, \bm W &\sim \mathcal{N}_p\left(\bm{G} \bm{\theta}_{t-1}, \bm W\right), \quad t=1,\ldots,T; \nonumber \\
\bm{\beta} &\sim \mathcal{N}_q(\bm{0}, 10 \mathbb{I}_q); \nonumber \\
\sigma^2 &\sim \text{InvGamma}(a_{\sigma^2}, b_{\sigma^2}), \quad a_{\sigma^2}=2, b_{\sigma^2} =0.1; \nonumber \\
\tau^2 &\sim \text{InvGamma}(a_{\tau^2}, b_{\tau^2}), \quad a_{\tau^2}=2, b_{\tau^2} =0.1; \nonumber\\
\diag(\bm W) &= (w_1,\ldots,w_p)';\, w_l  \sim \text{InvGamma}(a_{w_l}, b_{w_l}), \quad a_{w_l}=2, b_{w_l} =0.1, \quad l=1\ldots,p; \nonumber \\
\kappa &\sim U[0, 2\delta], \quad \delta \text{ is the maximum spatial distance}; \nonumber 
\end{align}
Whenever possible, we use conjugate priors for the hyperparameters to obtain closed-form full conditionals. Specifically, we assume an inverse Gamma (InvGamma) prior distribution with a small mean and relatively small variances for the variance hyperparameters, namely $\tau^2$, $\sigma^2$, and $w_l$ for $l=1,\ldots,p$. We achieve this by choosing fixed values for their corresponding shape and rate parameters of inverse Gamma distribution, denoted by $a$ and $b$, respectively. This results in closed-form full conditionals inverse gamma distributions for $\tau^2$, $\sigma^2$, and $w_l$ for $l=1,\ldots,p$; see Section~S1.6, S1.5, and S1.4, respectively, in Supplementary Material. 
The full conditional distribution of the range parameter $\kappa$ does not have a closed form, so we use a standard random walk Metropolis-Hastings algorithm to update these parameters; see Sections~S1.7 in Supplementary Material. By construction, the SPDE basis weight parameters $\bm R_t$ have closed-form full conditionals, but these conditionals depend on the previous and next steps of $\bm R_t$, specifically on $\bm R_{t-1}$ and $\bm R_{t+1}$, which makes parallel updates of $\bm R_t$ challenging. Forward filtering backward sampling \cite[FFBS,][]{carter1994gibbs} is a natural algorithm for updating the temporal components, but due to the large number of locations $n$, the computational cost per MCMC iteration is $\mathcal{O}(n^3 T)$, making it inefficient. Therefore, we update these components sequentially using Gibbs sampling, which is slightly computationally more expensive. However, due to the sparse precision matrices, the computation of full conditionals is simplified; see Section~S1.2 in the Supplementary Material for more details. Similarly, the full conditionals of the temporally evolving state vector $\bm\theta_t$ corresponding to temporal covariates have closed-form full conditionals but depend on the previous and next states of $\bm \theta$. While it is possible to perform Gibbs updates sequentially over time, FFBS algorithms are more accurate and known to provide faster convergence. Since the number of purely temporal covariates $p$, or harmonics is often small, it is feasible to use FFBS for updating $\bm \theta$, which is further simplified by our assumption of using purely temporal covariates; see Section~S1.3 in Supplementary Material. Finally, since the full conditional distribution of the rate $\lambda_t(\bm s)$ does not have a closed form, we update them using the preconditioned Metropolis-Adjusted Langevin Algorithm 
\citep[pMALA,][]{girolami2011riemann}. This results in faster convergence and does not require tuning, unlike the adaptation of the variance component used in standard MALA \citep{roberts2009examples}  algorithms.
\begin{algorithm}[t!]
\caption{Hybrid MCMC sampler for the sparse ST--DGLM of Section \ref{subsec:SPDEdetails}} 
\label{alg:mcmc-sampler-sparse}
\begin{algorithmic}[1]
  \State Initialize parameters $\bm \Theta =(\kappa^0, \sigma^{20}, \tau^{20}, \bm W^0, \bm \lambda^0, \bm R^0, \bm \theta^0)$ 
  \State initiate $j \to 0 $
  \While{$j < N^{mcmc}$} \Comment{$N^{mcmc}$ is total MCMC iterations}
  \State Propose $\kappa^*$ using normal random walk and accept or reject it using Metropolis correction. See the details of posteriors in Section~S1.7 in Supplementary Material. If accept, $\kappa^{j+1} = \kappa^*$ else $\kappa^{j+1} = \kappa^j$  
      \State Propose $\tau^{2\star}$, from its full conditional closed form distribution (Inverse gamma) of Section~S1.6 in Supplementary Material and accepts it with probability 1.
     \State Propose $\sigma^{2\star}$, from its full conditional closed form distribution (Inverse gamma) detailed in Section~S1.5 in Supplementary Material and accept it with probability 1.
     \For{$l$ in $1: p$}
      \State Propose $w_l^{\star}$ from its full conditional closed form distribution (Inverse gamma) detailed in Section~S1.4 in Supplementary Material and accepts it with probability 1.
       \EndFor
      \For{$t$ in $1: T$}
   \State Propose $\bm R_t^{\star}$ using the Gibbs algorithm detailed in Section~S1.2 in Supplementary Material
    \EndFor
   \For{$t$ in $1: T$}
    \For{$i$ in $1: n$}
   \State Propose $\lambda_{t,i}^\star$ using using the the pre-conditioned MALA algorithm detailed in Section~S1.1 in Supplementary Material \Comment{update it in vectorized fashion avoiding for-loops}
    \EndFor
    \EndFor
   \State Update $\bm \theta^{\star} = (\bm \theta_1^{\star}, \ldots, \bm \theta_T^{\star})'$ using the forward filtering backward sampling (FFBS) algorithm detailed in Section~S1.3 in Supplementary Material.
  \EndWhile
  \State Outputs: samples $\bm \Theta_1,\ldots,\bm \Theta_N$
\end{algorithmic}
\end{algorithm}

In Algorithm~\ref{alg:mcmc-sampler-sparse}, we showcase the detailed pseudocode of the hybrid MCMC sampler used to infer the sparse model of Section \ref{subsec:SPDEdetails}. The details for the most general ST--DGLM of Section \ref{subsec:dens-DGLM} are straightforward and can be obtained as a particular case of the  MCMC sampler detailed in Algorithm~\ref{alg:mcmc-sampler-sparse}. Specifically, the hierarchical model and the MCMC updates of all the model parameters corresponding to dense covariance matrices detailed in Section \ref{subsec:dens-DGLM} are the same as in Section~S1 of the Supplementary Material, but with the following replacements: $N=n$ means mesh nodes are exactly at the observed locations, meaning $\bm A = \mathbb{I}_n$; replace $\bm Q_\kappa^{-1}$ with $\bm\Omega$, where $\bm\Omega$ is a dense $n \times n$ Mat\'ern correlation matrix; and replace the notation $\bm R_t$ with $\bm \mu_t$. 

\section{Details of spatiotemporal predictions}
\setcounter{equation}{0}
\label{appd:appednB}
\subsection{Missing Value Imputation}
For missing value imputation, the target space-time points, denoted as $Y_t(\bm{s})$, lie within the observation window. These values, which may be assigned as \texttt{NA}'s in our MCMC sampler, may be filled using the posterior samples of the rate, $\lambda_t(\bm{s})$. Specifically, missing values may be imputed by first sampling $\lambda_t(\bm{s})$'s from their posterior distribution using the MCMC schemes detailed in Algorithm~\ref{alg:mcmc-sampler-sparse} for each missing time $t$ and spatial location $\bm{s}$ and then impute $Y_t(\bm{s})$ by sampling from
$$
Y_t(\bm{s}) \sim \text{Poisson}\left[\exp\left\{\lambda_t(\bm{s})\right\}\right].
$$
\subsection{Spatial interpolations or kriging}
\label{subsec:kriging-sparse}
In spatial interpolation or kriging, our goal is to predict the value of a spatial process at unobserved locations based on observed data. Our model setup allows for kriging through the partitioning of latent Gaussian process. The main task here involves specifying the predictive distributions 
$
\pi(\bm{\mu}_{\text{prd}} \mid \bm{\mu}, \cdot) \quad \text{and} \quad \pi(\bm{\theta}_{\text{prd}} \mid \bm{\theta}, \cdot)
$
in \eqref{eq:pred_predictor_level}, specifically for single space-time points $(\bm{s}_p, t)$ with the distributions 
$
\pi(\mu_t(\bm{s}_p) \mid \bm{\mu}, \cdot) \quad \text{and} \quad \pi(\bm{\theta}_{\text{prd}} \mid \bm{\theta}, \cdot).
$
In the case of spatial interpolation, since $\bm{\theta}$ is purely temporally varying, $\pi(\bm \theta_{\text{prd}} \mid \bm{\theta}, \cdot)$ is deterministic and may be set to 1 in \eqref{eq:pred_predictor_level}. 
Spatial interpolation of the component $\mu_t(\bm{s}_p)$ at space-time location $(\bm{s}_p, t)$ can be obtained using the partitioning scheme of Gaussian random vectors. Specifically, to perform spatial interpolation at a new location $\bm{s}_p$ using the model, we start by considering the model structure that describes the evolution of $\mu_t(\bm{s}_p)$ at each location $\bm{s}_p$ over time $t$. The increments $\omega_t(\bm{s})$'s are normally distributed with zero mean and Mat\'ern  covariance structure $\sigma^2 \bm{\Omega}$. Given observations at $\bm{s}_1, \ldots, \bm{s}_n$, our goal is to interpolate (predict) the process at a new location $\bm{s}_p$. To achieve this, we extend the Matérn correlation matrix to include the new location $\bm{s}_p$, resulting in an extended covariance matrix $\bm{\Omega}^*$ of size $(n+1) \times (n+1)$, expressed as:
\begin{align}
\bm{\Omega}^* = 
\begin{pmatrix}
\bm{\Omega} & \bm{\Upsilon}^* \\
(\bm{\Upsilon}^*)' & \Upsilon{**}
\end{pmatrix},
\end{align}
where $ \bm{\Upsilon}^* = \left( \Omega(\bm{s}_1, \bm{s}_p), \ldots, \Omega(\bm{s}_n, \bm{s}_p) \right)' $ represents the covariance between the new location and the observed locations, and $ \Upsilon{**} = \Omega(\bm{s}_p, \bm{s}_p) $ is the variance at the new location. Therefore the conditional distribution of $ \mu_t(\bm{s}_p) $ given $ \bm \mu_t = (\mu_t(\bm{s}_1), \ldots, \mu_t(\bm{s}_n))'$  is a multivariate normal distribution:
\[
\mu_t(\bm{s}_p) \mid \bm \mu_t \sim \mathcal{N} \left( \bm{\Upsilon}^{*'} \bm{\Omega}^{-1} \bm{\mu}_t, \sigma^2 \left( \Upsilon_{**} - \bm{\Upsilon}^{*'} \bm{\Omega}^{-1} \bm{\Upsilon}^* \right) \right).
\]
 To calculate the interpolated value at $ \bm{s}_p $, we consider the mean of $ \mu_t(\bm{s}_p) \mid \bm \mu_t $, i.e.;
$
\mathbb{E}[\mu_t(\bm{s}_p) \mid \bm{\mu}_t] = \bm{\Upsilon}^{*'} \bm{\Omega}^{-1} \bm{\mu}_t.
$
Therefore, the predicted log-intensity maybe written as:
\begin{align*}
\lambda_t(\bm{s}_p) = \bm{\Upsilon}^{*'} \bm{\Omega}^{-1} \bm{\mu}_t + \bm{F}_t' \bm{\theta}_t + \bm{X}_t(\bm{s}_p)'\bm{\beta} + \varepsilon_{t}(\bm{s}_p),
\end{align*}
where  ${\varepsilon}_t(\bm{s}_p)$ is the random error simulated from a normal distribution with zero mean and variance $\tau^{2}$, and $\bm X_t(\bm{s}_p)$ is the observed covariates vector at time $t$ and new spatial locations $\bm{s}_p$.

For the sparse ST-DGLM model detailed in Section \ref{subsec:SPDEdetails}, the log-rate at the new spatial locations $\bm{s}_p$ and time $t$  may be obtained solely from the information of $\bm{a}(\bm{s}_p)$, which projects the SPDE mesh node locations $\bm s_1^*,\ldots,\bm s_N^*$ to the new location of interest, $\bm{s}_p$. Therefore,
\begin{align*}
\lambda_t(\bm{s}_p) = \bm{a}(\bm{s}_p)' \bm{R}_t + \bm{F}_t' \bm{\theta}_t + \bm{X}_t(\bm{s}_p)' \bm{\beta} + \varepsilon_t(\bm{s}_p).
\end{align*}
Here, note that $\bm{R}_t$ remains unchanged across different spatial locations since the mesh node locations are fixed given the graphical domains which ensures that $\bm{R}_t$ is the same for all spatial locations, whether observed or predicted.

\subsection{Future Forecasting}
\label{subsec:future-forecast-sparse}
Future forecasting involves predicting the values of the process at future time $T+h, h=1,\ldots,$ by propagating the latent states forward in time. To do so, we need to specify the predictive distribution $\pi(\mu_{T+h}(\bm s) \mid \bm \mu, \cdot)$ and $\pi(\theta_{T+h} \mid \bm \theta, \cdot)$ in \eqref{eq:pred_predictor_level}.  To do this, we propagate the states $\bm{\mu}_{T+h}$ and $\bm{\theta}_{T+h}$ for $h = 1, 2, \ldots$, starting from $h = 0$ at the last observed time points. These states are sequentially updated for $h = 1, 2, \ldots$, with $h = 1$ representing the one-step-ahead future prediction, and so on.
Let $\bm{\mu}_{T}$, $\bm{\theta}_{T}$, $\sigma^{2}$, $\kappa$, and $\bm W$ be the posterior samples obtained using the MCMC sampler detailed in Appendix~\ref{appd:appednA}. Then we can simulate the posterior sample of $\mu$'s and $\theta$'s in future times by propagating the latent states forward in time as follows, for $h = 1, 2, \ldots$ :
\begin{align*}
 \bm{\mu}_{T+h} \mid \bm{\mu}_{T+h-1}, \sigma^{2}, \kappa \sim \mathcal{N}_n\left\{\bm{\mu}_{T+h-1}, \sigma^{2} \bm \Omega\right\}. \\   
 \bm{\theta}_{T+h} \mid \bm{\theta}_{T+h-1}, \bm{W} \sim \mathcal{N}_p\left(\bm{G} \bm{\theta}_{T+h-1}, \bm{W}\right)
\end{align*}
Therefore, we can obtain the forecasting log-rate  parameter by combining the propagated states to predict future values for site $\bm s$ (observed site) as
\begin{align*}
\lambda_{T+h}(\bm{s}) = {\mu}_{T+h} + \bm{F}_{T+h}' {\bm{\theta}}_{T+h} + \bm{X}_{T+h}(\bm{s})'{\bm{\beta}} + \varepsilon_{T+h}(\bm{s}), \quad, h=1,2,\ldots
\end{align*}
Similarly for the forecasting based on sparse ST-DGLM model we propagate $\bm{R}_{T+h}$, $h = 1, 2, \ldots$ as
\begin{align*}
\bm{R}_{T+h} \mid \bm{R}_{T+h-1}, \sigma^{2}, \kappa \sim \mathcal{N}_N\left\{\bm{R}_{T+h-1}, \sigma^{2} \bm{Q}_{\kappa}^{-1}\right\},  \quad, h=1,2,\ldots,
\end{align*}
and so the rate is
\begin{equation*}
\lambda_{T+h}(\bm{s}) = \bm{a}(\bm{s})' \bm{R}_{T+h} + \bm{F}_{T+h}' {\bm{\theta}}_{T+h} + \bm{X}_{T+h}(\bm{s})'{\bm{\beta}} + \varepsilon_{T+h}(\bm{s}).
\end{equation*}

\subsection{Spatiotemporal Predictions}
\label{subsec:spacetime-predictions-sparse}
Spatiotemporal predictions involve predicting the values of the process at both specific new spatial locations $\bm{s}_p$ and future time points $T+h, h=1, 2, \ldots$. This is done by explicitly combining spatial prediction (Section \ref{subsec:kriging-sparse}) and temporal prediction (Section \ref{subsec:future-forecast-sparse}). 
For the ST-DGLM model, the rate at future points $T+h, h=1, 2, \ldots$, and new locations $\bm{s}_p$ are given by:
\begin{align*}
\lambda_{T+h}(\bm{s}_p) = \bm{\Upsilon}^{*'} \bm{\Omega}^{-1} \bm{\mu}_{T+h} + \bm{F}_{T+h}' {\bm{\theta}}_{T+h} + \bm{X}_{T+h}(\bm{s}_p)'{\bm{\beta}} + \varepsilon_{T+h}(\bm{s}_p),
\end{align*}
where $\bm{\Upsilon}^{*'}$ and $\bm{\Omega}^{-1}$ are constructed as in Section \ref{subsec:kriging-sparse}, and $\bm{\mu}_{T+h}^*$ is the intercept at the new location $\bm{s}_p$ and future time points. At time point $T$, we predict the intercept at the new location as:
\begin{align*}
\mathbb{E}[\mu_T(\bm{s}_p) \mid \bm{\mu}_T] = \bm{\Upsilon}^{*'} \bm{\Omega}^{-1} \bm{\mu}_T,
\end{align*}
where $\bm{\mu}_T = (\mu_T(\bm{s}_1), \mu_T(\bm{s}_2), \ldots, \mu_T(\bm{s}_n))'$ represents the observed intercept at time $T$.
Then, we obtain $\mu_{T+h}(\bm{s}_p)$ by propagating as follows:
\begin{align*}
\mu_{T+h}(\bm{s}_p) \mid \mu_{T+h-1}(\bm{s}_p) \sim \mathcal{N} \left(\mu_{T+h-1}(\bm{s}_p), \sigma^2 \left( \Upsilon_{**} - \bm{\Upsilon}^{*'} \bm{\Omega}^{-1} \bm{\Upsilon}^* \right) \right),
\end{align*}
where $\Upsilon_{**} = \Omega(\bm{s}_p, \bm{s}_p)$ is the variance at the new location.

For spatiotemporal prediction based on sparse ST--DGLM models at future time $T+h$, the spatial predictions ${\lambda}_{T+h}(\bm{s}_p)$ are combined with the temporal evolution of the states as
\begin{align*}
\lambda_{T+h}(\bm{s}_p) = \bm{a}(\bm{s}_p)' \bm{R}_{T+h} + \bm{F}_{T+h}' {\bm{\theta}}_{T+h} + \bm{X}_{T+h}(\bm{s}_p)'{\bm{\beta}} + \varepsilon_{T+h}(\bm{s}_p),
\end{align*}
where $\bm{R}_{T+h}$ and $\bm{\theta}_{T+h}$ are constructed similarly as in Section \ref{subsec:future-forecast-sparse}.

\section{Detail of the synthetic data experiment}
\label{appd:appednC}
\setcounter{equation}{0}
While simulating synthetic data from the ST--DGLM in Section~\ref{sec:synthetic_exp} of the  manuscript, the log-rate parameter is defined as 
\begin{align} 
\label{eq:intensity-exp2_appnd}
   \lambda_t(\bm{s}) &=  \mu_t(\bm{s}) + \bm{F}_t' \bm{\theta}_t + \bm{X}_t(\bm{s})' \bm{\beta} + \varepsilon_{t}(\bm{s}),
\end{align}
where:
\begin{itemize}
    \item $\bm{\theta}_t = (\theta_{1t}^{\text{ws}_1}, \theta_{2t}^{\text{ws}_1})'$ is a $2 \times 1$ vector of purely dynamic coefficients.
    \item $\bm{F}_t(\bm{s}_i) = (1, 0)'$ is a $2 \times 1$ dimensional vector at the space-time location $(t, \bm{s})$, and the corresponding evolution matrix is given by
    \[
    \bm{G}_{t} = \bm G =  \begin{pmatrix}
            \cos\left(\frac{2\pi}{7}\right) &  \sin\left(\frac{2\pi}{7}\right) \\
           -\sin\left(\frac{2\pi}{7}\right) &  \cos\left(\frac{2\pi}{7}\right)
        \end{pmatrix}.
    \]
    \item $\bm{X}_t(\bm{s}_i) = (X_t(\bm{s}_i)^1, X_t(\bm{s}_i)^2, X_t(\bm{s}_i)^3)'$ includes three space-time covariates simulated randomly from $\mathcal{N}(0, 1)$.
\end{itemize}

\section{Details about the models used for data application}
\setcounter{equation}{0}
\label{appd:appednD} 
\subsection{Model details}
\begin{enumerate}
    \item 
\textbf{Baseline model: Bayesian Poisson GLMs} 
 \begin{align*}
 Y_t(\bm{s}) \mid \lambda_t(\bm{s}) &\stackrel{\rm{ind}}{\sim} \mathrm{Poisson}\left[\exp\{\lambda_t(\bm{s})\}\right],\quad \bm s\in \mathcal{S} \subset \mathbb{R}^2, t=1,\ldots,T,  \\
\lambda_t(\bm{s}) &= \mu + \bm{X}_t(\bm{s})'\bm{\beta} + \varepsilon_{t}(\bm{s}).
\end{align*}
\item \textbf{BKTR model} of \citep{lei2024scalable, lei2024scalable}, which is an efficient spatiotemporally varying coefficient regression model: 
\begin{align*}
    \tilde{Y}_t(\bm{s}) &= \log\{Y_t(\bm{s})\}, \\
    \tilde{y}_t(\bm{s}) &= \bm{X}_t(\bm{s})' \bm{\beta}_t(\bm{s}) + \varepsilon_t(\bm{s}),
\end{align*}
where the $q$ coefficients $\bm{\beta}_t(\bm{s})$ are approximated by low-rank tensor factorization \citep{bahadori2014fast}, bypassing heavy computational difficulties.
 \item \textbf{ST--DGLM} is fitted using the following components:
 \begin{align*}
\lambda_t(\bm{s}) = \mu_t(\bm s) + \bm{F}_t' \bm{\theta}_t + \bm{X}_t(\bm{s})'\bm{\beta} + \varepsilon_{t}(\bm{s}),
\end{align*}
where $\boldsymbol{F_t} = (1,0,1,0)$ and the corresponding $\bm G$ matrix is:
{\scriptsize
\begin{align*}
 \bm G = \begin{pmatrix}
 \cos(2 \pi \frac{1}{7}) & \sin(2 \pi \frac{1}{7}) & 0 & 0   \\
 -\sin(2 \pi \frac{1}{7}) & \cos(2 \pi \frac{1}{7}) & 0 & 0  \\ 
 0 & 0 & \cos(2 \pi \frac{2}{7}) & \sin(2 \pi \frac{2}{7})   \\
 0 & 0 & -\sin(2 \pi \frac{2}{7}) & \cos(2 \pi \frac{2}{7}) \\ 
\end{pmatrix}
\end{align*}
}
 \item \textbf{Sparse ST--DGLM model:}
in this model, all components remain the same as in the ST--DGLM, except for the intercept part $\mu_t(\bm{s})$, which is now replaced by $\mu_t(\bm{s}) = \bm{a}(\bm{s})'\bm{R}_t$. Consequently, the log-rate parameter is given by 
\begin{align*}
\lambda_t(\bm{s}) = \bm{a}(\bm{s})'\bm{R}_t + \bm{F}_t' \bm{\theta}_t + \bm{X}_t(\bm{s})'\bm{\beta} + \varepsilon_t(\bm{s}).
\end{align*}
\item \textbf{BayesNF model}:
in this model \citep[BayesNF;][]{saad2024scalable}, observations $Y_t(\bm{s})$ are modeled as:
\begin{align*}
Y_t(\bm{s}) \sim \text{Dist}\{g(F_t(\bm{s})), \bm\gamma\},
\end{align*}
where the latent process $F_t(\bm{s})$ is defined as:
\begin{align*}
F_t(\bm{s}) = h(\boldsymbol{X}_t(\bm{s});\bm \beta) + \eta_t(\bm{s}).
\end{align*}
The covariates $X_t(\bm{s})$ are defined to include multiple components:
\begin{align*}
\boldsymbol{X}_t(\bm{s}) = \text{Linear terms} + \text{spatiotemporal interaction terms} + \text{Fourier features}.
\end{align*}
Temporal Fourier features capture periodic or seasonal patterns, while spatial Fourier features address both low-and high-frequency spatial variations, enhancing the model's ability to represent complex spatial structures. Additional exogenous covariates, such as temperature, visibility, or population density, can also be included. The systematic mean $h(\boldsymbol{ X}_t(\bm{s}); \bm \beta)$ is modeled through a neural network as:
\begin{align*}
h(\boldsymbol{X}_t(\bm{s}); \bm \beta) = \text{NN Layers}(\text{Activations}(\boldsymbol{X}_t(\bm{s}); \bm \beta)),
\end{align*}
where nonlinear activation functions, such as ReLU, ELU, or Tanh, enable the network to learn complex relationships.  
 \end{enumerate}
\subsection{Specification of the covariates for all the models}
The covariate vector $\boldsymbol{X}_t(\bm{s})$ varies depending on the model under consideration. For the \textbf{Poisson GLM} and \textbf{BKTR} models, the covariate vector is defined as 
\begin{align*}
\boldsymbol{X}_t(\bm{s}) = \left(T_t, v_t, w_t, pd_t, wd_t, yr_t, \sin\left(\frac{2\pi t}{7}\right), \cos\left(\frac{2\pi t}{7}\right), ele(\bm{s}), ws(\bm{s}), pp(\bm{s})\right)',
\end{align*}
where $T_t$, $v_t$, $w_t$, $pd_t$, $wd_t$, and $yr_t$ represent temperature, visibility, wind speed, precipitation amount (mm), a weekend/weekday dummy variable, and the year index, respectively. The purely spatial covariates $ele(\bm{s})$, $pp(\bm{s})$, and $ws(\bm{s})$ correspond to elevation, population density, and walk scores, respectively. These covariates collectively capture temporal, spatial, and spatiotemporal variations essential for modeling in each framework.

In contrast, for the ST--DGLM, \textbf{sparse} ST--DGLM models, and BayesNF the covariate vector is specified as 
\begin{align*}
\boldsymbol{X}_t(\bm{s}) = \left(v_t, w_t, pd_t, wd_t, yr_t, ele(\bm{s}), ws(\bm{s}), pp(\bm{s})\right)'.
\end{align*}

\end{document}

%% file: macros.tex
\def\frac#1#2{{\textstyle{#1\over#2}}}

\def\diag{{\rm diag}}

\DeclareSymbolFont{AMSb}{U}{msb}{m}{n}
\DeclareMathSymbol{\Natural}{\mathbin}{AMSb}{"4E}
\DeclareMathSymbol{\Integer}{\mathbin}{AMSb}{"5A}
\DeclareMathSymbol{\Real}{\mathbin}{AMSb}{"52}
\DeclareMathSymbol{\Rational}{\mathbin}{AMSb}{"51}
\DeclareMathSymbol{\Imaginary}{\mathbin}{AMSb}{"49}
\DeclareMathSymbol{\Complex}{\mathbin}{AMSb}{"43} 
\DeclareMathSymbol{\Disk}{\mathbin}{AMSb}{"44} 
\def\bi{\begin{itemize}}
\def\ei{\end{itemize}}
\def\bd{\begin{description}}
\def\ed{\end{description}}
\def\ben{\begin{enumerate}}
\def\een{\end{enumerate}}
\def\bv{\begin{verbatim}}
\def\ev{\end{verbatim}}

\def\hat#1{{\widehat{#1}}}

\def\2to{{\ {\buildrel 2\over \longrightarrow}\ }}

\def\I1ton{{$I_1,\ldots,I_n$}}
\def\X1ton{{$X_1,\ldots,X_n$}}
\def\Y1ton{{$Y_1,\ldots,Y_n$}}
\def\Z1ton{{$Z_1,\ldots,Z_n$}}
\def\R1ton{{$R_1,\ldots,R_n$}}
\def\e1ton{{$e_1,\ldots,e_n$}}
\def\t1ton{{$t_1,\ldots,t_n$}}
\def\x1ton{{$x_1,\ldots,x_n$}}
\def\y1ton{{$y_1,\ldots,y_n$}}
\def\z1ton{{$z_1,\ldots,z_n$}}